\documentclass[preprint,11pt]{elsarticle}
\usepackage[top=1in, bottom=1in, left=1in, right=1in]{geometry}

\usepackage{graphicx}
\usepackage{amssymb}

\usepackage{lineno}
\usepackage{comment}
\usepackage{amsmath}
\usepackage{tabu}
\usepackage{xcolor}
\usepackage{setspace}
\usepackage{bm}
\usepackage[colorlinks=true,bookmarksopen,pdfsubject={algorithms},linkcolor={blue}, anchorcolor={black}, citecolor={blue}, filecolor={magenta}, menucolor={black}, pagecolor={red},backref=none,urlcolor={blue}]{hyperref}

\usepackage{natbib}
\setcitestyle{square, comma, numbers,sort&compress, super}

\DeclareMathOperator*{\argminA}{arg\,min}

\journal{Computational Materials Science}

\begin{document}
\begin{spacing}{1.15}
\begin{frontmatter}





\title{Monte Carlo simulation of order-disorder transition in refractory high entropy alloys: a data-driven approach  \footnote{\footnotesize{This manuscript has been co-authored by UT-Battelle, LLC, under contract DE-AC05-00OR22725 with the US Department of Energy (DOE). The US government retains and the publisher, by accepting the article for publication, acknowledges that the US government retains a nonexclusive, paid-up, irrevocable, worldwide license to publish or reproduce the published form of this manuscript, or allow others to do so, for US government purposes. DOE will provide public access to these results of federally sponsored research in accordance with the DOE Public Access Plan (http://energy.gov/downloads/doe-public-access-plan).}}}

\author[First]{Xianglin Liu}

\author[Second]{Jiaxin Zhang\corref{cor1}}
\ead{zhangj@ornl.gov}

\author[Third]{Junqi Yin}
\author[Fourth,Fivth]{Sirui Bi}
\author[Third]{Markus Eisenbach}
\author[Six]{Yang Wang}

\cortext[cor1]{Corresponding author}

\address[First]{ Materials Science and Technology Division, Oak Ridge National Laboratory, Oak Ridge, TN 37830, USA}
\address[Second]{Computer Science and Mathematics Division, Oak Ridge National Laboratory, Oak Ridge, TN 37830, USA}
\address[Third]{Center for Computational Sciences, Oak Ridge National Laboratory, Oak Ridge, TN 37830, USA}
\address[Fourth]{Computational Science and Engineering Division, Oak Ridge National Laboratory, Oak Ridge, TN 37830, USA}
\address[Fivth]{Department of Civil and Systems Engineering, Johns Hopkins University, Baltimore, MD 21210, USA}
\address[Six]{Pittsburgh Supercomputing Center, Carnegie Mellon University,  Pittsburgh, PA 15213, USA}


\begin{abstract}
High entropy alloys (HEAs) are a series of novel materials that demonstrate many exceptional mechanical properties. To understand the origin of these attractive properties, it is important to investigate the thermodynamics and elucidate the evolution of various chemical phases. In this work, we introduce a data-driven approach to construct the effective Hamiltonian and study the thermodynamics of HEAs through canonical Monte Carlo simulation. The main characteristic of our method is to use pairwise interactions between atoms as features and systematically improve the representativeness of the dataset using samples from Monte Carlo simulation. We find this method produces highly robust and accurate effective Hamiltonians that give less than 0.1 mRy test error for all the three refractory HEAs: MoNbTaW, MoNbTaVW, and MoNbTaTiW. Using replica exchange to speed up the MC simulation, we calculated the specific heats and short-range order parameters in a wide range of temperatures. For all the studied materials, we find there are two major order-disorder transitions occurring respectively at $T_1$ and $T_2$, where $T_1$ is near room temperature but $T_2$ is much higher. We further demonstrate that the transition at $T_1$ is caused by W and Nb while the one at $T_2$ is caused by the other elements. By comparing with experiments, {\color{black} the results provide insight into the role of chemical ordering in the strength and ductility of HEAs.}

\end{abstract}



\end{frontmatter}


\section{Introduction}
\label{S:1}
High entropy alloys (HEAs) \cite{ADEM:ADEM200300567, CANTOR2004213,ye2016high} are a class of novel materials that attracts significant interest due to their superior mechanical properties \cite{Gludovatz1153, NatureComNiCoCr, SENKOV2011698, Fueaat8712, senkov_miracle_chaput_couzinie_2018}. Compared to conventional alloys, the most distinctive feature of HEAs is that they are composed of multiple principal elements in approximately equal proportions. The random mixing of principal elements substantially enhances the configurational entropy, which can overcome the enthalpy of mixing to stabilize the random solid solution phase. Although initially, the completely random phase was the focus of people's attention, recently there is a growing interest to further improve the mechanical properties of HEAs through the tuning of the secondary phase, nano-structure, and short-range order (SRO) \cite{Fueaat8712, GWALANI2017254, RAZA201897, ROGAL2020108716, Ding8919}. For example, second-phase particles rich in Cr and V \cite{El-Atwanieaav2002} is believed to be the origin of the excellent radiation tolerance in a W-based refractory HEAs, and the dislocation pinning by $1.3\pm 0.4$ nm Ar bubbles is found to be responsible for the high mechanical damage tolerance in MoNbTaW thin films \cite{TUNES2019107692}. Moreover, compared to the ordered phase, the high chemical complexity of HEAs introduces strong chemical fluctuation to the random phase \cite{YOSHIDA2019201, ZHANG2019424, liu2019chemical}, which has a profound effect on quantities governing the plastic deformation, such as stacking fault energy, dislocation core structure, and Peierls potential, as revealed in recent works \cite{VARVENNE2016164,  SMITH2016352, e20090655, Ding8919, NatureCommShijun, LIU2019107955}. As a result, it is reasonable to expect that the occurrence of order-disorder transition in HEAs can have a significant effect on the mechanical properties. To investigate these effects, it is highly desirable to have an accurate, efficient, and versatile tool for the study of thermodynamics in HEAs.

Density functional theory (DFT) is a powerful method to calculate material properties from first principles. Combined with Monte Carlo (MC) methods, DFT is widely applied to study finite temperature systems \cite{widom_2018, Eisenbach_2019, Fernandez-Caballero2017}. One simple approach to combine DFT with MC is to use DFT to evaluate the energy at each Monte Carlo step, which is obviously computationally very expensive due to the large number of MC steps. For example, using a supercell of 250-atom \cite{PhysRevB.93.024203}, the ``brute-force" statistical simulation of the order-disorder transition in CuZn alloy requires the calculation of 600,000 DFT energies of different chemical configurations, which is only practical on supercomputers with more than 100,000 CPU cores. A more common approach is to build an energy model (effective Hamiltonian) from DFT calculations, and employ the computationally cheap lattice model in the Monte Carlo simulation. Two different approaches are commonly utilized to construct the energy model. The first one is to calculate the effective cluster interactions (ECI) directly from DFT using the linear response theory of concentration waves. The implementations of this approach are typically based on the coherent potential approximation (CPA), such as the $S^{(2)}$ theory \cite{PhysRevLett.50.374}, the generalized perturbation method (GPM) \cite{0305-4608-6-11-005}, and the embedded-cluster method (ECM) \cite{PhysRevB.36.4630}. While these methods can provide important physical insight by elegantly connecting the order parameters with the electronic band structure \cite{mu_pei_liu_stocks_2018}, their analytical nature and heavy dependence on CPA renders them difficult to implement for complex alloys. The second approach is to extract the ECI parameters from the energies of different chemical configurations. The best-known example of this strategy is the cluster expansion method \cite{PhysRev.81.988, SANCHEZ1984334, NPJ_Widom, PhysRevB.72.165113, doi:10.1021/ja9105623, chang2019clease, alidoust2020density}, where the {\color{black}ECI} parameters are typically calculated with the Connolly and Williams approach (also known as the structure inversion method) \cite{PhysRevB.27.5169}. In principle, cluster expansion provides a complete basis to represent different chemical configurations. In practice, a truncation has to be applied to retain the dominant cluster terms \cite{PhysRevLett.92.255702} for the calculation to be feasible. In contrast to the aforementioned CPA methods, in cluster expansion, DFT only plays the role of calculating energy data. The separation of data generation and model fitting makes cluster expansion easy to be combined with various DFT implementations, and be widely used for the study of thermodynamics in materials.  

The application of cluster expansion in HEAs, however, faces additional challenges due to the rapid increase of the number of parameters in the model. First, a simple structure inversion is no longer reliable to determine a large number of model parameters due to the bias-variance trade-off \cite{MEHTA20191}, which means a complex model is prone to overfitting. To build a robust energy model, a large and representative dataset is required, and techniques such as regularization, cross-validation, and model selection, need to be applied. Second, since in practice only pair and triplet interactions within a few atomic shells are typically included in the cluster expansion of HEAs, the included clusters in the model should not be assumed to form a complete basis. Due to the above two points, it is helpful to treat this problem from the perspective of machine learning \cite{ML_Nature2018, Korman_npj, 2020NPJ_Li}, where cluster expansion can be viewed as a tool for providing ``physics-inspired" features, and machine learning techniques can be employed to optimize the model. This idea is similar to the previous efforts in the cluster expansion community, such as Bayesian cluster expansion \cite{PhysRevB.80.024103, PhysRevB.81.012104}, optimal selection of structures \cite{seko2009cluster}, and renormalized interactions \cite{PhysRevB.99.134206}. Nevertheless, it is also important to note that, the features in the model are not necessarily limited by cluster expansion, nor the model has to be linear.
One should also note that cluster expansion seeks to obtain a model Hamiltonian for all chemical concentrations, and such a requirement typically compromises the accuracy of the energy model. In this work we focus on canonical MC simulation, where the chemical composition is fixed. Therefore, by ``configuration", we refer to the arrangement of atoms in the lattice,  under the constraint of fixed chemical composition.

One recent trend is to utilize machine learning approaches to construct an efficient data-driven model, predict the mechanical or chemical properties, and improve the performance via material design \cite{yu2020new, wu2020uncovering, hu2018two} in multicomponent systems \cite{pei2020machine, huang2019machine,jafary2019applying, li2019machine, zhang2020phase, zhou2019machine,dai2020theoretical}. In principle, if a large dataset is provided, methods such as deep neural networks (DNN) can be employed for this purpose. In practice, a few challenges need to be addressed in the implementation of the machine learning approach. First, machine learning methods are generally data-hungry, therefore it is essential to generate the data efficiently via high-fidelity simulations. However, most DFT implementations are computationally expensive due to the intrinsic cubic scaling behavior. Improving the DFT calculation speed is particularly critical for HEAs due to the need for a relatively large supercell to represent various chemical phases. Second, the dataset needs to be representative to cover the whole high-dimensional configuration space. Otherwise, the learned model only performs well in the training data but shows fake predictability in the ``unseen" testing dataset. This means a statistical sampling scheme needs to be devised so that the chemical configurations in the dataset have various degrees of order and disorder, and contain states in the whole relevant energy range. Both challenges need to be addressed for the investigation of the thermodynamics of multicomponent systems, e.g. HEAs, through Monte Carlo simulation.

In this work, we develop a data-driven framework to address the above challenges. To speed up the data generation process, the linear-scaling self-consistent multiple scattering (LSMS) method \cite{PhysRevLett.75.2867} is employed to calculate the configurational energy from first principles. The dataset is first initialized with random configurations to obtain a preliminary energy model, from which a canonical MC simulation is carried out and a small configuration sample is drawn. This sample of configurations are calculated with LSMS, and these newly obtained data points are then added into the original dataset to improve the data representativeness. Using the improved dataset, we perform an adaptive model construction and achieve a significant improvement in model accuracy. The updated model is employed to conduct the study of thermodynamics through Monte Carlo simulation.  Note that the direct calculation of the DFT energies of the MC samples is a key feature of our method. It not only enables us to evaluate the model accuracy in large MC supercell, but also renders it straightforward to implement an adaptive, incremental learning policy. By using the same supercell in both DFT calculation and MC simulation, the two procedures are linked together to form a self-consistent loop, as distinct from other methods \cite{Korman_npj}. Using this proposed approach, the specific heats and short-range order parameters of three refractory HEAs, MoNbTaW, MoNbTaVW, and MoNbTaTiW are evaluated to study the order-disorder transitions.

\section{Methods}

\subsection{Energy model}

\begin{figure}[!ht]    
    \centering
\includegraphics[width=0.4\textwidth]{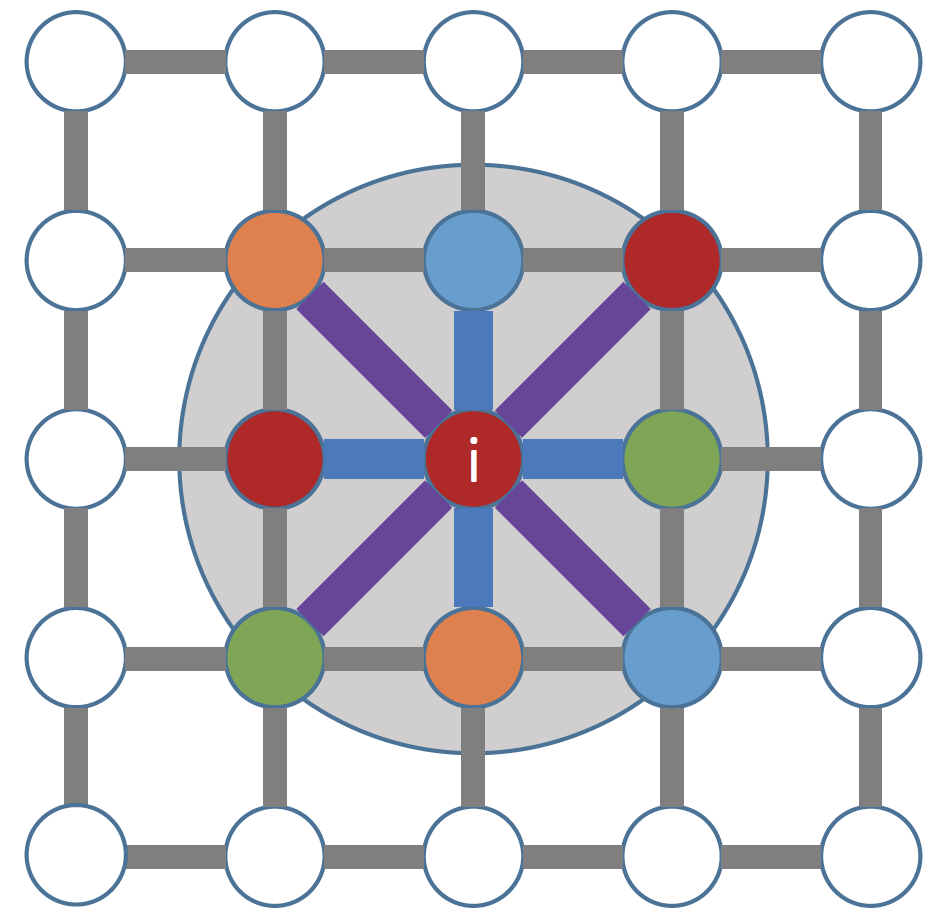}
  \caption{ (Color online) A schematic to illustrate the dependence of the local energy $E_i$ on the local chemical environment. The different colors of the atoms signify different chemical species. The effective pair interactions of the nearest neighbor and next nearest neighbor are shown explicitly.} 
\label{fig:EPI_schematic}
\end{figure}

{\color{black}The total energy of a system can be expressed as a summation of the local atomic energies, which can be approximated with an effective Hamiltonian.} Conventional cluster expansion is difficult for HEAs due to the large number of multi-site interactions, thus here we adopt the effective pair interactions (EPI) model \cite{liu2019machine}. In the EPI model, the effective Hamiltonian of the system is made up of the chemical pair interactions of the centering atom with neighboring atoms within some cut-off radius, as illustrated in Fig.~\ref{fig:EPI_schematic}. The local chemical environment is specified by $\vec{\sigma} = (\sigma^0, \sigma^1, \cdots, \sigma^{N_{n}-1})$, which denotes the chemical species of the $N_n$ neighboring atoms. The local energy $E_i$ is given by 
\begin{align}
E_i = \sum_{f} V^f \pi^f(\vec{\sigma}_i)  + V^p_i + V^0 + \epsilon,
\label{Linear_EPI}
\end{align}
where $\epsilon$ is the uncertainty of the EPI model, $V^0$ is the bias term same for all sites, $V_i^p$ is a single-site term depending only on the chemical component $p$ of atom $i$, $V^f$ are the EPI parameters, and $\pi^f$ are the number of pair interactions of type $f$. The feature index $f$ is actually made up of three parts $(p, p', m)$, representing the element of the local atom, the element of the neighboring atoms, and the coordination shell, respectively. For a canonical system, summing up the local energies over all sites, the total energy is then given by
\begin{align}
    {\color{black}E \approx N \sum_{p'<p,m} V_m^{pp'} \Pi_m^{pp'} + \mathrm{const} + \epsilon,}
\label{Total_EPI}
\end{align}
where $N$ is the total number of atoms and $\Pi_{m}^{pp'}$ is the proportion of $pp'$ interaction in the $m$-th neighboring shell. Note that due to the fixed chemical composition in canonical system, the single-site term $V_i^p$ has been absorbed into the constant, and the number of independent EPI parameters is $M(M-1)/2$ for a $M$-component system, which is the reason for the $p'<p$ requirement. {\color{black} The EPI model has demonstrate high accuracy to approximate the total energy in recent studies \cite{liu2019machine,ZHANG2020108247}. For fixed chemical concentrations, the pairwise iteration plays a dominant role in determining the Hamiltonian but higher-order terms may have limited impact on the EPI model.}

\subsection{{\color{black}EPI parameter estimate}}

The EPI parameters can be determined by solving the linear system given by Eq.~\eqref{Total_EPI}, which can be cast in a matrix form 
\begin{equation}
    \mathbf{E} = \mathbf{V}\mathbf{\Pi} + \mathbf{\epsilon}
\end{equation}
where $\mathbf{E}$ is a column vector in which the $i$-th element in the total energy $E_i$, $\mathbf{\Pi}$ is a matrix containing the proportion of $pp'$ interaction in the $m$-th neighboring shell, and $\mathbf{V}$ is a column vector in which the $m$-th neighboring shell is $V_m$ which also includes $M(M-1)/2$ elements. Determining the EPI model in Eq.~\eqref{Total_EPI} is equivalent to finding the parameter vector $\mathbf{V}$ that minimizes the residual sum of squared  (RSS) errors $|| \mathbf{E} - \mathbf{V \Pi} ||_2^2$.  
A typical solution is the ordinary least squares (OLS) method, which performs well in the overdetermined systems. {\color{black} However, a large number of parameters can lead OLS to fit the noise in the data \cite{chang2019clease}, which compromises the predictability of the model. The OLS method is also sensitive to outliers and thus leads to unstable and unreliable estimates. }

Ridge regularization has been proved to be an effective approach to avoid overfitting by adding an $\ell_2$ norm term to the cost function, which amounts to adding a Gaussian prior to the EPI parameters. In this context, the optimal EPI parameters $\mathbf{V}^*$ can be solved by 
\begin{equation}
    \mathbf{V}^* = \argminA_{\mathbf{V}} || \mathbf{E} - \mathbf{V \Pi} ||_2^2 + \lambda || \mathbf{V}||_2^2, \label{eq: l2norm} 
\end{equation}
One problem in solving Eq.~\eqref{eq: l2norm} is how to optimally determine the $\ell_2$ regularization parameter $\lambda$. Rather than choosing the regularization parameter $\lambda$ by hand, Bayesian ridge regression treats $\lambda$ as a random variable and determines it automatically along with other model parameters by maximizing the log marginal likelihood. Consequently, the $\ell_2$ regularization in Eq. \eqref{eq: l2norm} is equivalent to finding a maximum a posterior (MAP) estimation \cite{zhang2019efficient,zhang2020quantification} given a Gaussian prior over $\bm{J}$ with precision $\xi^{-1}$, that is $p(\mathbf{V} | \xi )=\mathcal{N}(\bm{J} | 0, \xi ^{-1})$. Typically, a MAP estimation of the posterior distribution is obtained by Markov Chain Monte Carlo (MCMC) algorithm \cite{zhang2018quantification}, which is often computationally intensive and difficult to converge for high dimensional problem \cite{zhang2018effect}. In this work, we consider a conjugate prior for which the posterior distribution can be derived analytically and utilize the implementation in the scikit-learn package \cite{scikit-learn} for all the reported results. More hyperparameters details can be found in \cite{ZHANG2020108247}.  

{\color{black} Compared to the OLS method, Bayesian ridge regression has three advantages: 1) $\ell_2$ regularization can mitigate the overfitting issue and stabilize the estimator, and 2) Bayesian techniques can include regularization parameters in the estimation procedure so that the regularization parameter is not set in a hard sense but tuned to the data at hand and 3) Bayesian ridge regression estimates a probabilistic model of the regression model and is well-suite to take uncertainty (e.g. noise term) into consideration. }


\subsection{Monte Carlo simulation }
The thermodynamic observables are estimated using canonical Monte Carlo simulation, where both the temperature and the number of particles of each component are fixed. Specifically, each Monte Carlo update is made up of proposing a swap of chemical species between neighboring atoms and accepting it according to a Markov Chain updating scheme, which is the Metropolis algorithm in our case. After the system reaches equilibrium, the configurations follow Boltzmann distribution and samples can be drawn to calculate observables. Due to the competing interactions in the system, the replica exchange Monte Carlo (REMC) algorithm \cite{Swd86} is employed to speed up the simulation. In REMC, multiple simulations are performed simultaneously at different temperatures, and the systems exchange configurations according to a Metropolis-like probability that satisfies the detailed balance condition. The transition probability from a configuration $\{\vec{\sigma}\}_m$ simulated at temperature $T_m$ to a configuration $\{\vec{\sigma}\}_n$ simulated at temperature $T_n$ is
\begin{equation}
W(\{\vec{\sigma}\}_m,T_m|\{\vec{\sigma}\}_n,T_n)= \mathrm{min}\left[1,\exp(-\triangle)\right],
\end{equation}
where
\begin{equation}
\triangle=(1/k_BT_n-1/k_BT_m)({E}_m-{E}_n),
\end{equation}
and $k_B$ is the Boltzmann constant. Since the specific heats are non-divergent at the transition point for all the HEAs studied, it is optimal to set the simulation temperature in geometric series \cite{Kof04}. The Monte Carlo step hence consists of an atom swap trial for every lattice site, followed by a replica exchange update between neighboring temperatures. In practice, 5 different initial configurations are utilized to estimate the statistical error and ensure the equilibrium state is reached. During the Monte Carlo simulation, $10^6$ warm-up steps are discarded and the subsequent $10^7$ steps are used to estimate thermodynamic observables. In this work, the thermodynamic quantity we focus on are the specific heat and the short-range order parameter. The specific heat is calculated by evaluating the standard deviation of the energy. The Warren-Cowley short-range order parameter \cite{PhysRev.77.669, OWEN2016155} $\alpha_m^{pp'}$ is defined as
\begin{equation}
\alpha_m^{pp'} = 1 - \frac{P^{p|p'}_m}{c_p}, \label{SRO}
\end{equation}
where $c_p$ is the concentration of element $p$, and $P^{p|p'}_m$ is the probability of finding element $p$ at the $m$-th neighbor shell for a given element $p'$, and is calculated by averaging over the Monte Carlo samples.

\subsection{DFT calculation}
In the data generating process, larger supercells are needed to simulate the complex chemical environment, and a big number of chemical configurations are required to generate enough data. Both the two requirements substantially increase the computational cost, making DFT calculation the bottleneck of the calculation speed, therefore an efficient implementation of DFT is highly desirable. As mentioned in the introduction, the speed improvement is accomplished by the linear-scaling LSMS method. The LSMS method is a real space implementation of the Korringa-Kohn-Rostoker (KKR) method \cite{KORRINGA1947392, PhysRev.94.1111}. Its linear-scaling behavior is achieved by restricting the quantum scattering of the electrons within the so-called local interaction zone (LIZ). Note that all the electrostatic interactions are still explicitly calculated, therefore LSMS can reliably predict the small energy difference between different chemical configurations.

In practice, the dataset is initially made up of configurations generated randomly using supercells of different sizes. This is a simple approach to include various degrees of order and disorder in the configuration sample, because small supercells naturally give rise to ordered structures due to periodic boundary conditions, while large supercell configurations will be close to the random state. After determining the DFT energies of the configurations with LSMS, the data are split into training and test datasets. For all the three refractory HEAs, the lattice constants are chosen as 6.2 Bohr, the angular momentum cutoff is 3, the Barth-Hedin local-density approximation is used as the exchange-correlation functional, and the size of the LIZ is 59 atoms. To properly treat the heavier elements in the system, the scalar-relativistic equation is solved rather than the conventional Schr{\"o}dinger equation. In our experience, LSMS takes about 1 hour to calculate the energy of a 1000-atom system using 200 CPU cores ($\approx$ 5 nodes) in the Summit supercomputer, which means the calculation can also be accomplished with regular computing clusters. If needed, the excellent linear-scaling behavior of LSMS can be utilized to investigate systems of more than $10^5$ atoms. 

{\color{black}
\subsection{Improve data representativeness}
An additional benefit of the LSMS method is that it allows the use of the same supercell during data generation and Monte Carlo simulation, in contrast to the conventional practice of using different supercells in DFT and MC simulation. As a result, the configuration obtained from Monte Carlo simulation can be directly fed into LSMS to calculate the DFT energy. This allows the evaluation of the model for configurations unable to be described with small supercells, such as nano-precipitates. Moreover, by unifying the data generation and simulation process, the initial dataset can be improved through choosing a small set of MC configurations to be calculated with DFT. Depending on the goal of the simulation, different schemes can be devised to choose the MC configurations. For example, if one only wants to find the ground state, then a weighted sampling favoring low energy states can be adopted. In our case, since we are mainly interested in the temperature dependence of the observables, we adopt a simple scheme of mixing the configurations from a range of different temperatures up to $T_{max}$. The value of $T_{max}$ is set to be a little higher than the order-disorder transition temperature so that the ordered states not well represented in the initial random dataset can be incorporated. }

\subsection{Methodology workflow}

\begin{figure}[!ht]    
    \centering
\includegraphics[width=0.5\textwidth]{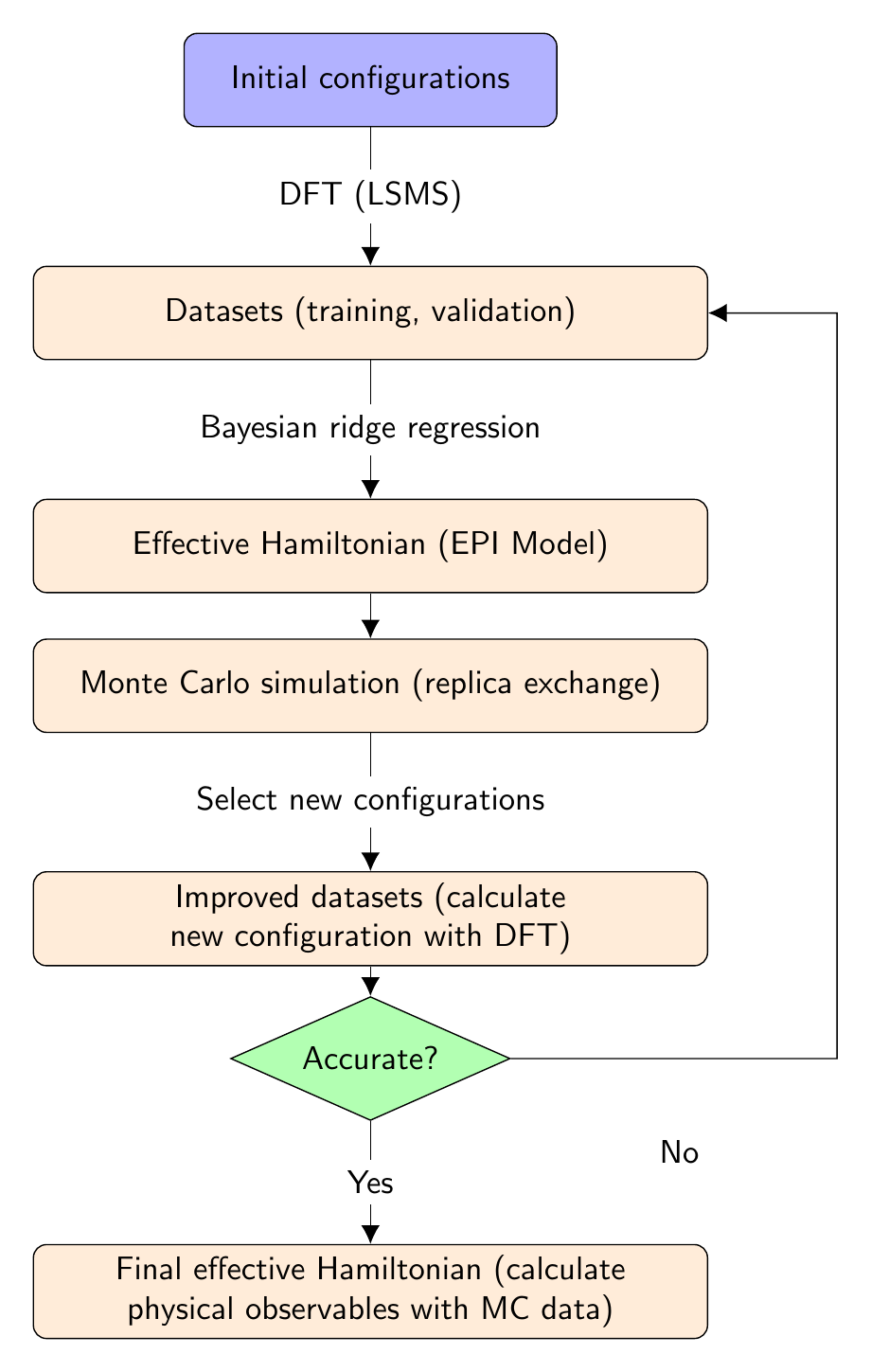}
  \caption{ {\color{black} (Color online) A diagram of the workflow of our method.} }
\label{fig:work_flow}
\end{figure}

The workflow of our method is shown in Fig.~\ref{fig:work_flow} and summarized as follows: 

\begin{itemize}
    \item Step 1: Initial dataset: Generate random configurations with different size supercells and calculate their energies with LSMS, using the same calculation parameters. Combine these DFT data of different supercells to include both ordered and disordered states. We denote the initial (current) dataset as $D$.  
    
    \item Step 2: Effective Hamiltonian: Split the data $D$ into training dataset $D_t$ and validation dataset $D_v$. Determine the EPI model parameters with Bayesian ridge regression. A model selection process is also carried out to optimize the hyper-parameters \cite{ZHANG2020108247}. For the EPI model, this corresponds to determine the coordination shell cutoff $m$.
    
    \item Step 3: Monte Carlo simulation: Carry out MC simulation using the learned data-driven energy model. Replica exchange is utilized to speed up the calculation. In this work, we use $10\times 10 \times 10$ supercell for demonstration.
    
    {\color{black}
    \item Step 4: New data points: Select a set of new configurations from the Monte Carlo samples to improve the data representativeness and increase the model accuracy. Calculate the DFT energies $\hat{E}$ of these new data points $D_n$ with LSMS, and compare them with $E$ calculated by the current EPI model. If the difference $\varepsilon = |\hat{E}-E|$ is less than a threshold $\hat{\varepsilon }$, the accuracy criterion is accepted, then go to step 5; otherwise, add these new data $D_n$ into the current dataset $D$  to update the energy model (go back to step 2). Note that the choice of $\varepsilon$ is somewhat case by case since it depends on the specific material. For the purpose of canonical MC simulation, $\varepsilon = 0.1$ mRy is generally good since it corresponds to a temperature difference of about 15 K.}
    
    \item Step 5: Physical observables: Calculate thermodynamic observables from the MC samples. In this work we are mainly interested in the SRO parameters and specific heats at different temperatures, in order to investigate the order-disorder transitions in the refractory HEAs. 

\end{itemize}


\section{Results}

\subsection{Accuracy of the effective Hamiltonian}
\begin{figure}[!ht]    
    \centering
\includegraphics[width=0.32\textwidth]{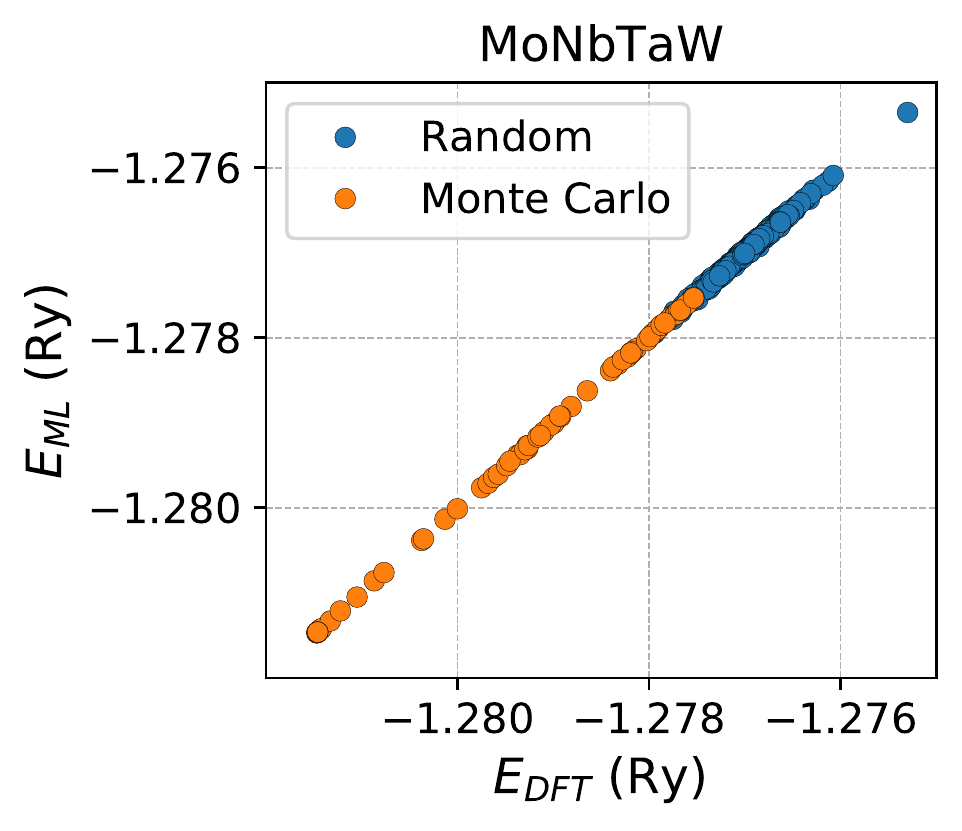}
\includegraphics[width=0.32\textwidth]{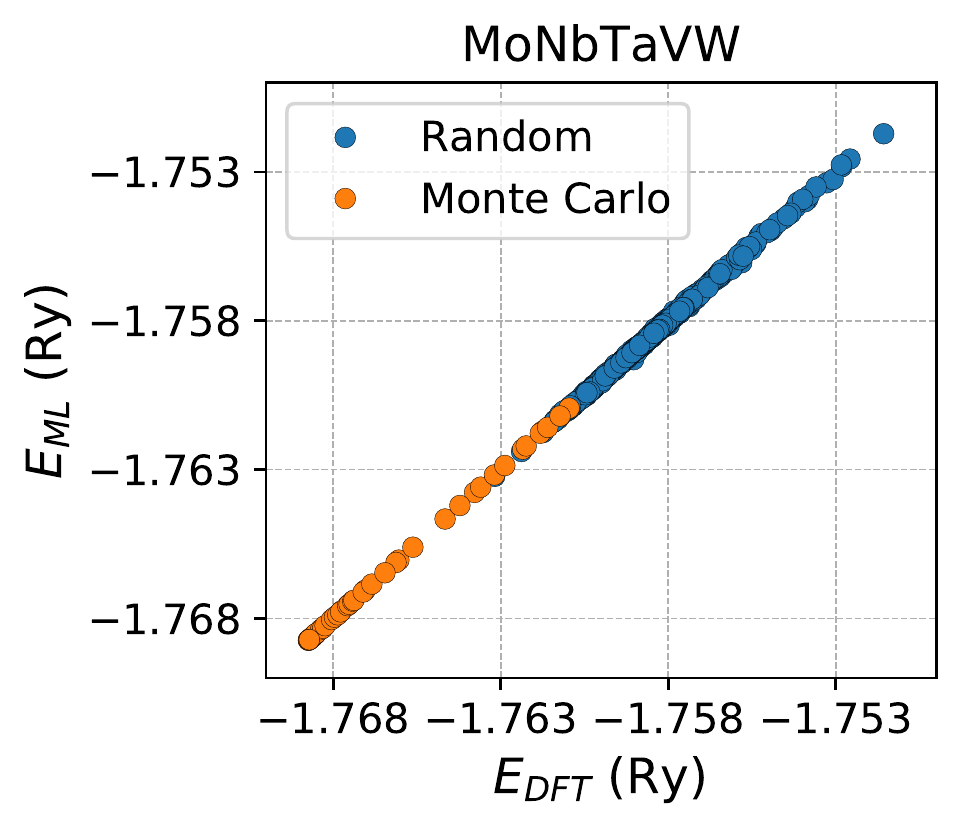}
\includegraphics[width=0.32\textwidth]{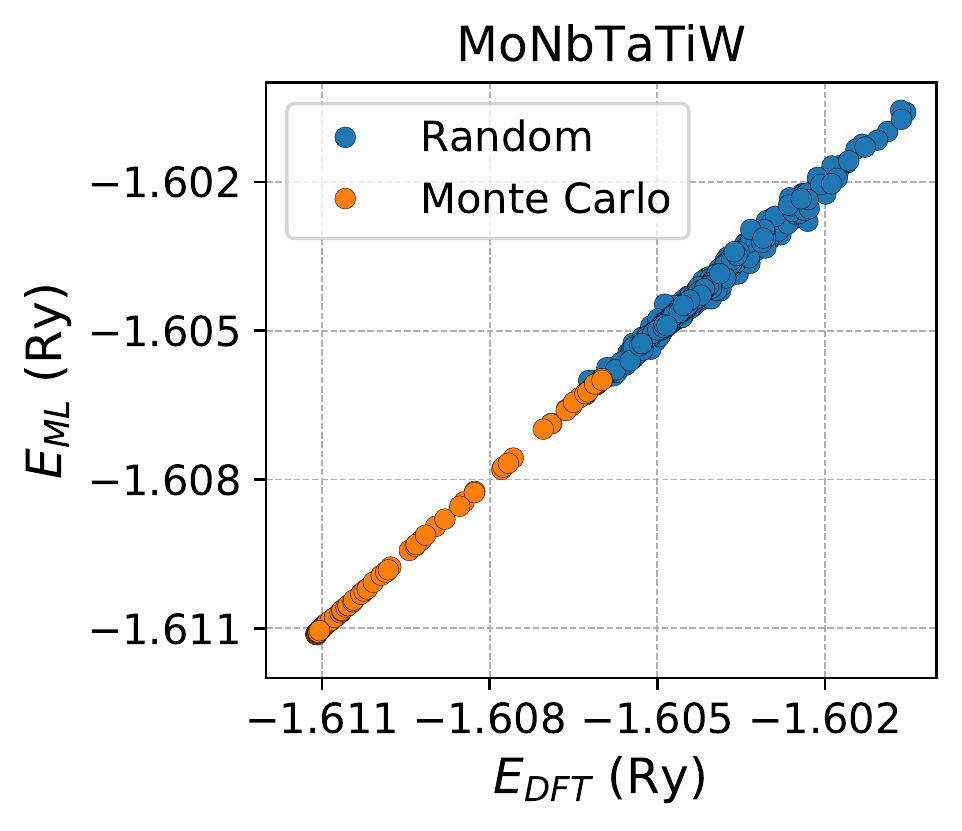}
  \caption{ (Color online) Comparison of the DFT energis against the energies predicted by the model for the three HEAs. The Blue circles represent the data points of random configurations using different supercells. The red dots represent the data points from Monte Carlo simulation.} 
\label{fig:energies}
\end{figure}

The energies of configurations generated randomly (random data) and from Monte Carlo simulations (MC data) are shown as x-axis data in Fig.~\ref{fig:energies}. For MoNbTaW, there are 72 MC data and a total of 704 random data from supercells of 64, 128, 256, and 512 atoms. For MoNbTaVW, there are 48 MC data and a total of 1232 random data from supercells of 20, 40, 80, 160, 320, 640, and 1280 atoms. For MoNbTaTiW, there are 85 MC data and 899 random data from supercells of 20, 40, 80, 160, and 320 atoms. From Fig.~\ref{fig:energies}, it is easy to see that while the random configurations (blue dots) well represent the states at elevated temperatures, they do not take into account the low temperature states. On the other hand, the low energy states are well incorporated in the Monte Carlo samples (orange dots). To evaluate the accuracy of the model, the dataset is split randomly into two halves, with one used for training and the other one for testing. From these data, A 6-shell EPI model is used to predict the DFT energies, and the results are displayed as the y-axis data in Fig.~\ref{fig:energies}. The reason to use 6 coordination shells in the model is detailed in \cite{ZHANG2020108247}, where the Bayesian information criterion (BIC) is employed to identify the best number of coordination shells. From Fig.~\ref{fig:energies}, it can be seen that this model gives a very accurate prediction of the DFT energy for all the three HEAs. In practice, we also tried to include higher-order interactions, for example, quadratic interaction terms to the energy model but did not see an improvement of accuracy. Based on the Occam's razor principle, we adopt the relatively simple EPI model throughout this work.

To illustrate the accuracy of the model in more quantitative detail, the calculated $R^2$ scores and root mean square errors (RMSEs) are listed in Tab.~\ref{tab:t2}. It can be seen that all the $R^2$ scores are higher than 0.99 and all the RMSEs are less than 0.1 mRy. To be specific, the test RMSEs are 0.0197, 0.0443, and 0.0992 for MoNbTaW, MoNbTaVW, and MoNbTaTiW, respectively. Moreover, the differences between the training and testing results are very small, indicating that the models are well trained and not being overfitted.

\begin{table}[!ht] 
\centering
\caption{Summary of the testing $R^2$ scores and root mean square errors (RMSEs) for the training and testing. The data include both the random configurations and the Monte Carlo samples. Half of the data is used for training and the other half for testing.}
\label{tab:t2}
\begin{tabular}{ccccc }
\hline
Material & $R^2_{\mathrm{train}}$ score & $R^2_{\mathrm{test}}$ score & $\mathrm{RMSE_{train}}$ (mRy) & $\mathrm{RMSE_{test}}$ (mRy) \\ \hline
MoNbTaW &0.9995 &0.9995 &0.0170 &0.0197   \\
MoNbTaVW &0.9997 &0.9994 &0.0322 &0.0443 \\
MoNbTaTiW &0.9978 &0.9968 &0.0805 &0.0992 \\ \hline
\end{tabular}
\end{table}

\begin{figure}[!ht]    
    \centering
\includegraphics[width=0.8\textwidth]{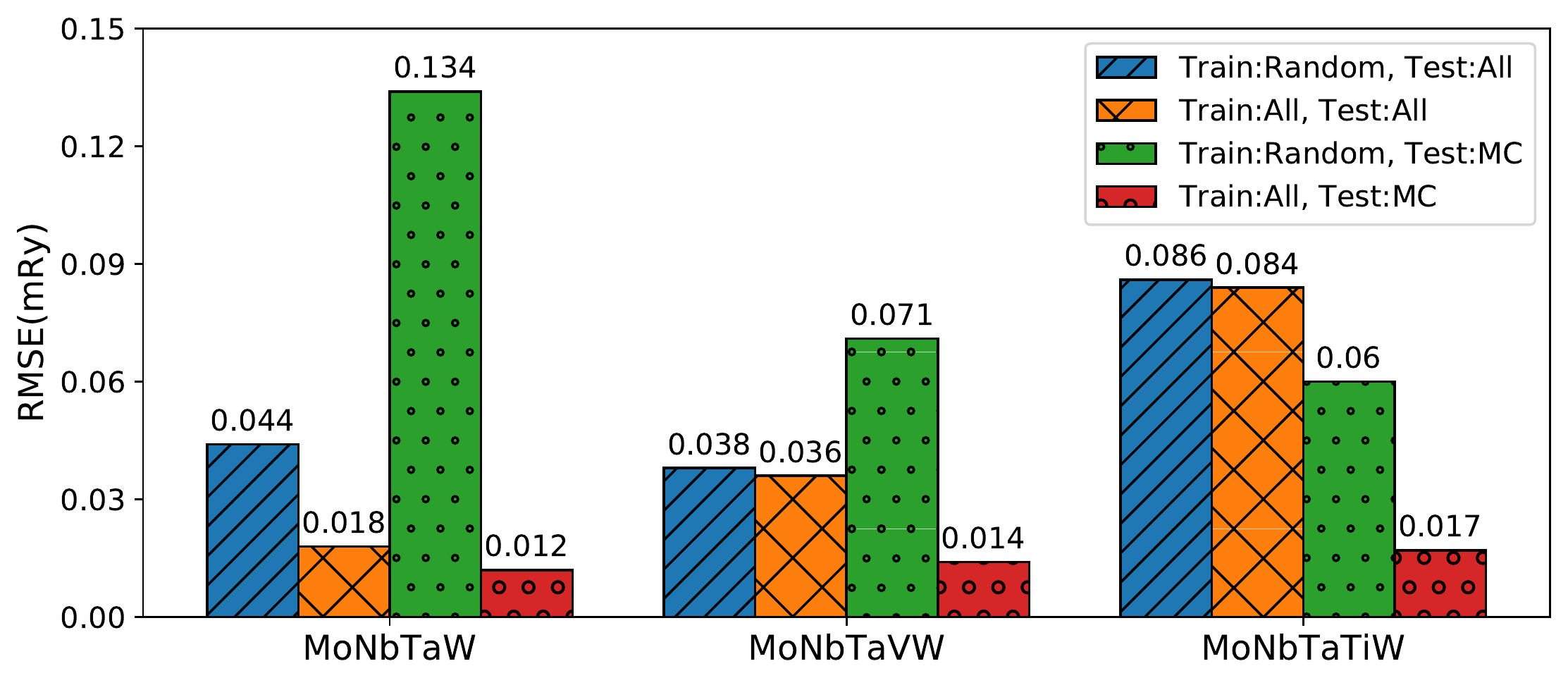}
  \caption{ (Color online) The root mean square errors (RMSEs) calculated with different training and testing datasets. In the plot legend, ``Random" represents using random configurations, ``MC" signifies using only Monte Carlo samples, and ``All" represents using both random configurations and Monte Carlo samples.} 
\label{fig:RMSE_bar}
\end{figure}

To demonstrate the effects of adding the MC data, two EPI models are trained separately with different datasets: one contains only the random data and the other one includes both the random data and MC data. The two models are then tested separately with the complete dataset and the MC dataset. The results are shown in Fig.~\ref{fig:RMSE_bar}. First, it is easy to observe that even by training with the random data, the obtained model still performs reasonably good for the MC samples, which is mostly outside the energy range of the random data. This demonstrates the robustness of the EPI model due to the use of the ``physics-informed" features.
Second, it can be seen that including the MC samples indeed reduces the RMSEs. The improvement is less significant when tested with the complete data. This is because the number of MC data is much less than the random one. When tested only with the MC data, the RMSEs are decreased substantially by almost one order of magnitude. These results underpin the importance of adding the MC data, particularly for the accurate prediction of low energy states. Finally, it is necessary to note that the red bar in Fig.~\ref{fig:RMSE_bar} is lower than the others, indicating that the EPI model is better at describing the low temperature ordered states than the random states. 

\subsection{ Investigation of effective pair interaction parameters}\label{EPI}
\begin{figure}[!ht]    
    \centering
\includegraphics[width=0.6\textwidth]{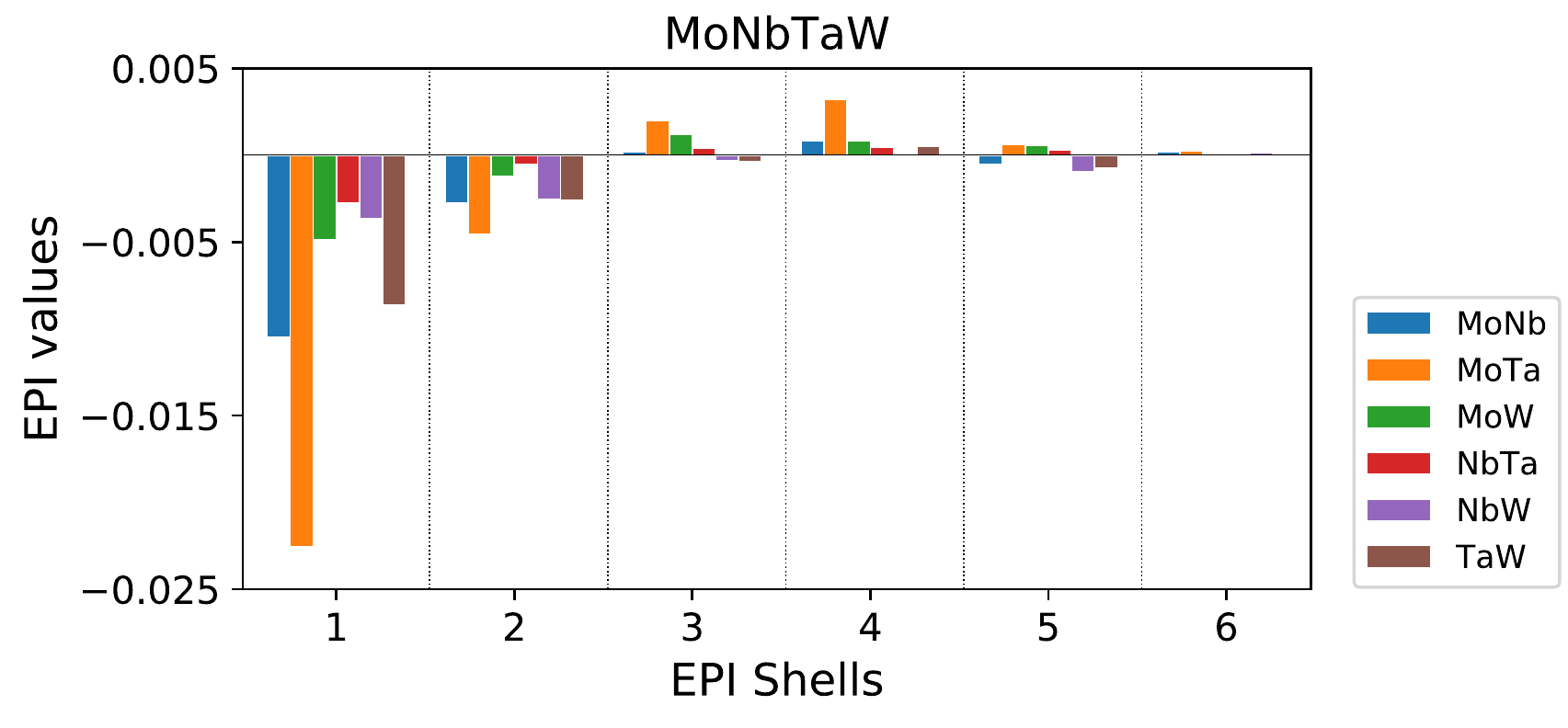}
\includegraphics[width=0.6\textwidth]{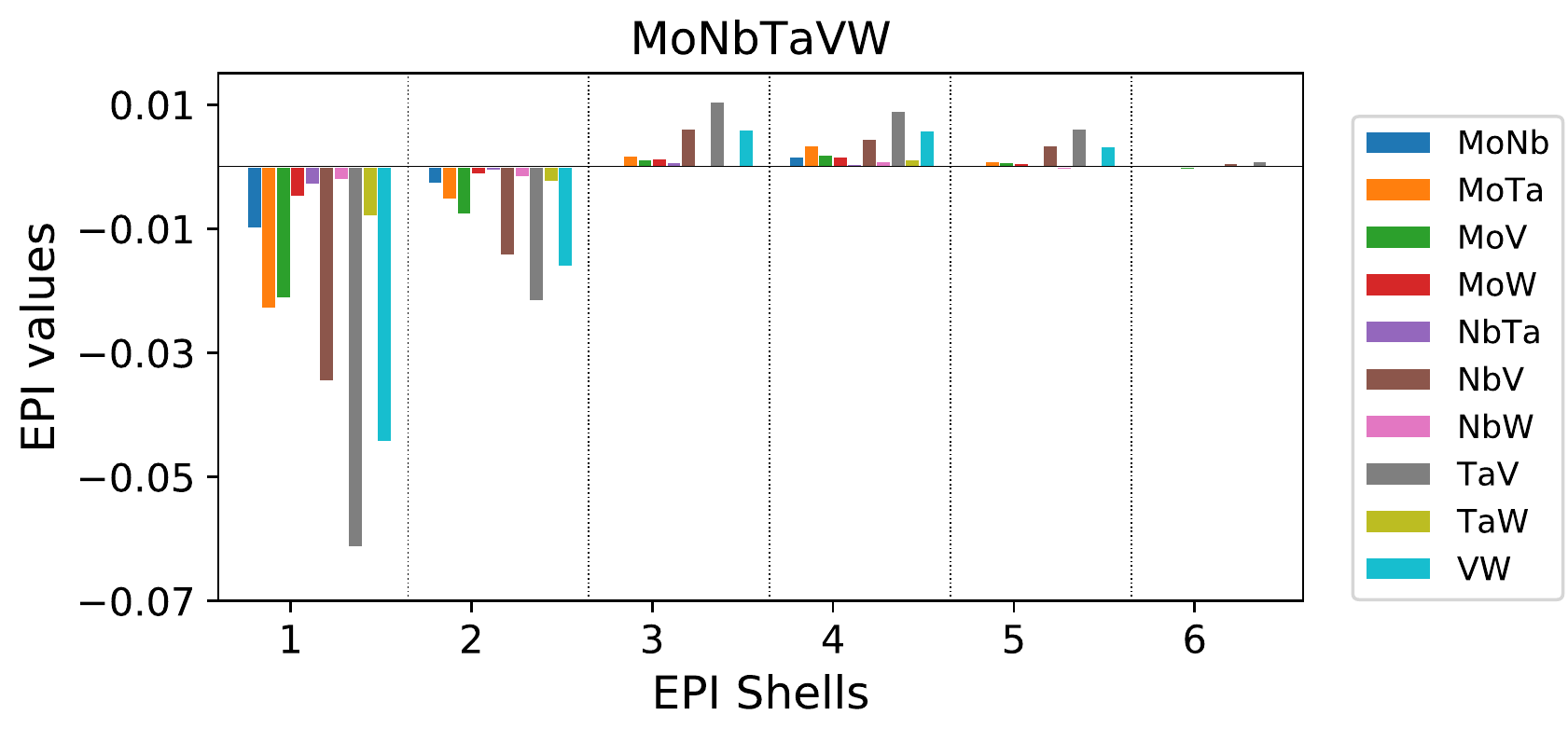}
\includegraphics[width=0.6\textwidth]{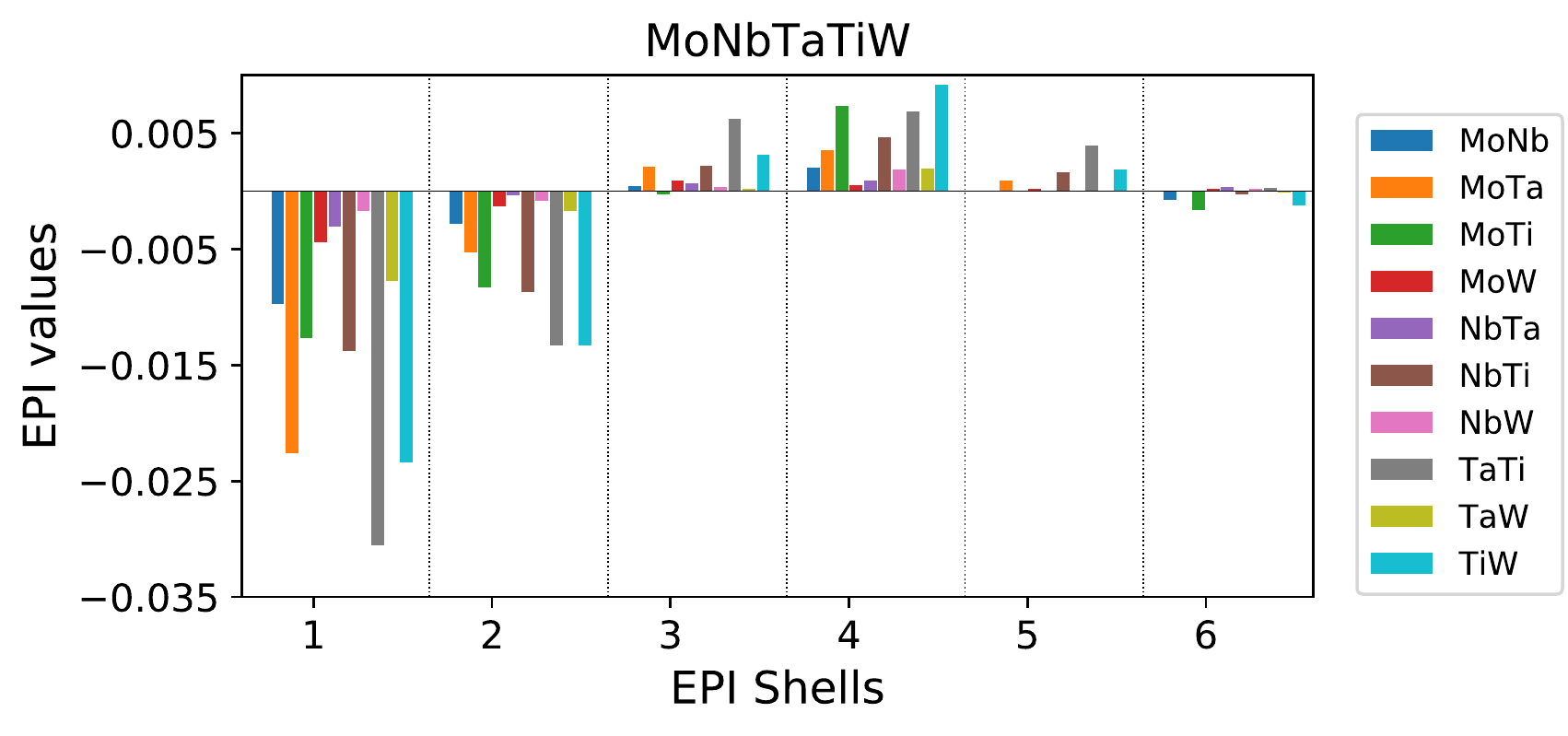}
  \caption{ (Color online) The EPI parameters of the three refractory HEAs.} 
\label{fig:EPI}
\end{figure}

To understand the microscopic origin of the strength and ductility of materials, it is important to study the atomic interactions, which corresponds to the EPI parameters in our model. The bonding profile between atoms affects the formation and movement of dislocations, therefore it plays an important role in the plastic deformation of materials. The element-element pair interactions \cite{YOSHIDA2019201} in HEAs are particularly interesting due to their complex chemical compositions. Generally speaking, if the interactions between different element pairs are of similar magnitudes, then it is easy for the atoms to change positions in the lattice since the energy difference is small. As a result, it is easier for this type of HEA to form random solid solutions. On the other hand, if the interactions are highly heterogeneous, then the HEA tends to form an intermetallic compound. Furthermore, it can be expected that HEAs with homogeneous pair interactions tend to demonstrate better ductility because the easy movement of atoms facilitates dislocation glide. 

The EPI parameters for the three refractory HEAs are shown in Fig.~\ref{fig:EPI}. First, it is easy to note that the nearest and next-nearest neighbor pair interactions are the dominant ones. Among the three HEAs, the pair interactions beyond the second neighbor are relatively weak in MoNbTaW, as compared to MoNbTaVW and MoNbTaTiW, which contain more frustrated long-range interactions. To be more specific, for MoNbTaW, the strongest interaction occurs among the MoTa pairs,  which is in agreement with the results in literature \cite{Huhn2013, ZHANG2020108247}. For MoNbTaVW, the result shows that V has much stronger interactions with other elements, with the TaV pair as the strongest. For MoNbTaTiW, the strongest interaction occurs in the TaTi pairs. From these results, we can observe that the strongest pair interactions occur among atoms with the largest difference of electronegativity. On the other hand, for all the materials, NbTa and NbW are the weakest pair interactions.

By comparing the EPI parameters of the three HEAs, a few predictions can be made about their properties, based on our previous arguments. First, the order-disorder transition temperature of MoNbTaVW should be much higher than MoNbTaW and MoNbTaTiW, because the addition of V introduces strong pair interactions to the material, including VW, TaV, and NbV, which hinder the formation of random phase. Second, the ductility of MoNbTaVW should be worse than the other two materials due to its heterogeneous element-element pair interactions. This is actually in agreement with the experimental observation that adding V decreases the ductility  \cite{senkov_miracle_chaput_couzinie_2018}. To draw a quantitative conclusion, a detailed study of the Peierls potentials of each material would be required.

\subsection{Order-disorder transition}

\begin{figure}[!ht]    
    \centering
\includegraphics[width=0.8\textwidth]{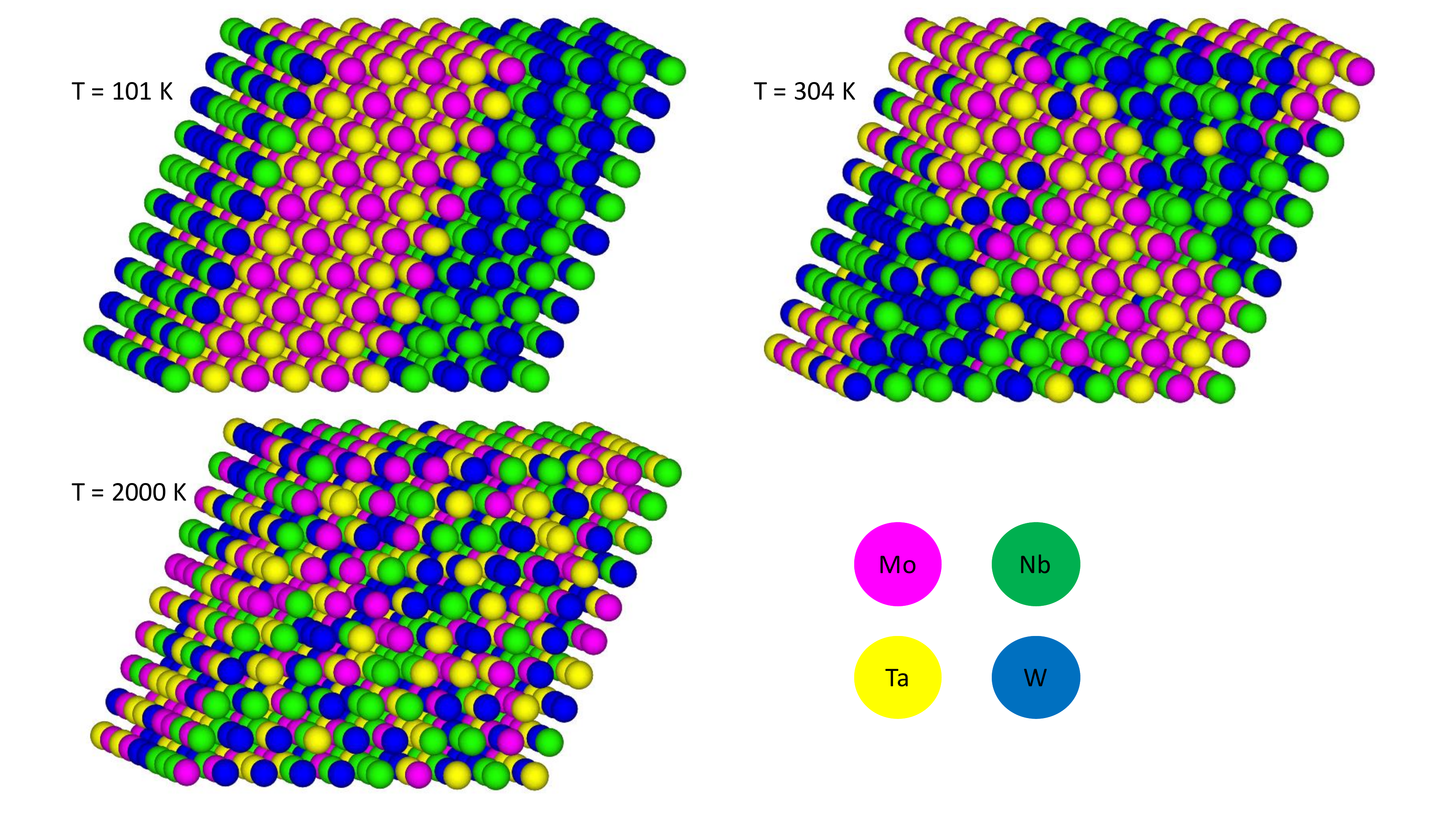}
  \caption{ (Color online) Snapshots of the structure of MoNbTaW at 101 K, 304 K, and 2000 K.} 
\label{fig:snap}
\end{figure}

\begin{figure}[!ht]    
    \centering
\includegraphics[width=0.45\textwidth]{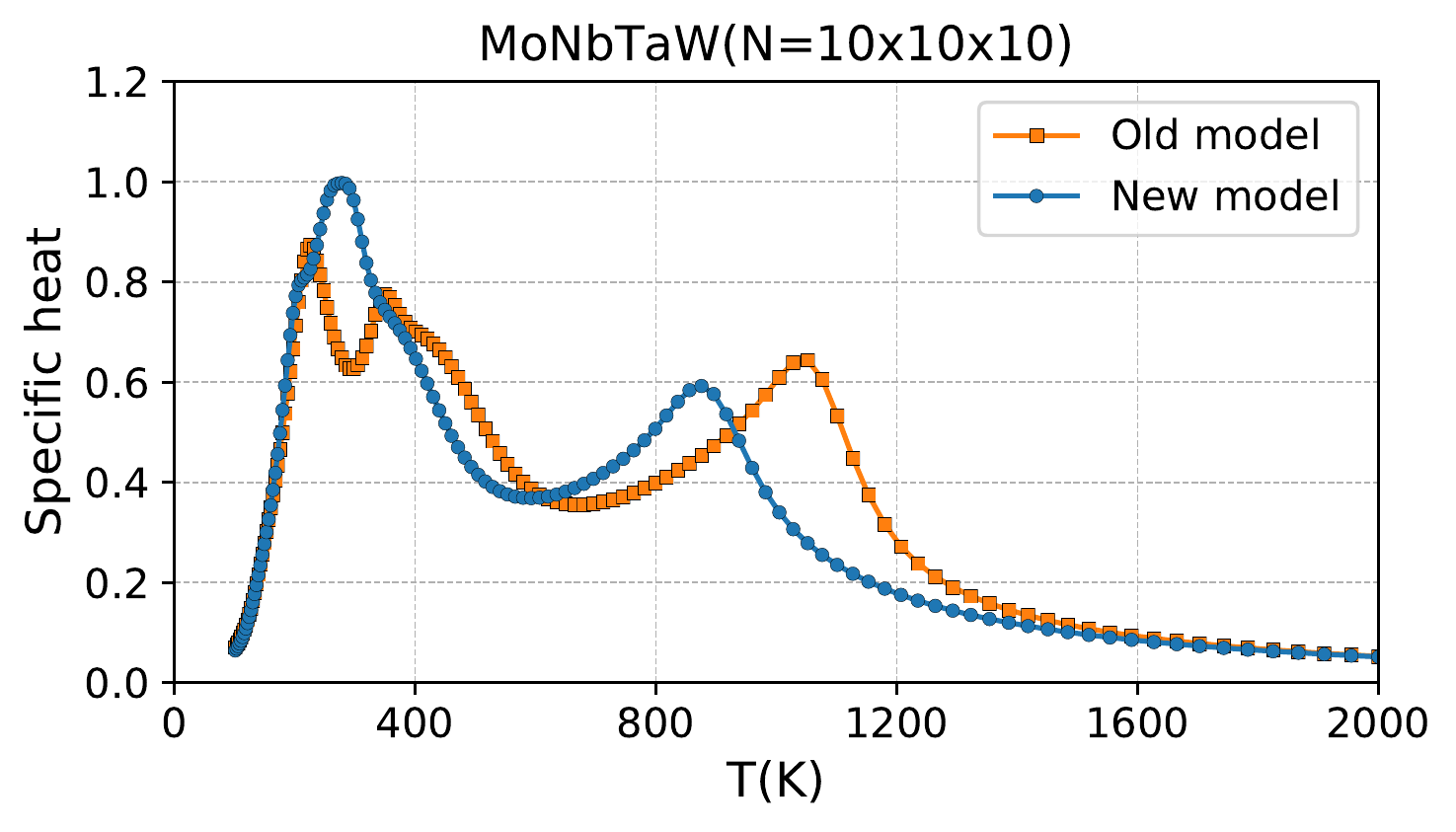} \\
\includegraphics[width=0.45\textwidth]{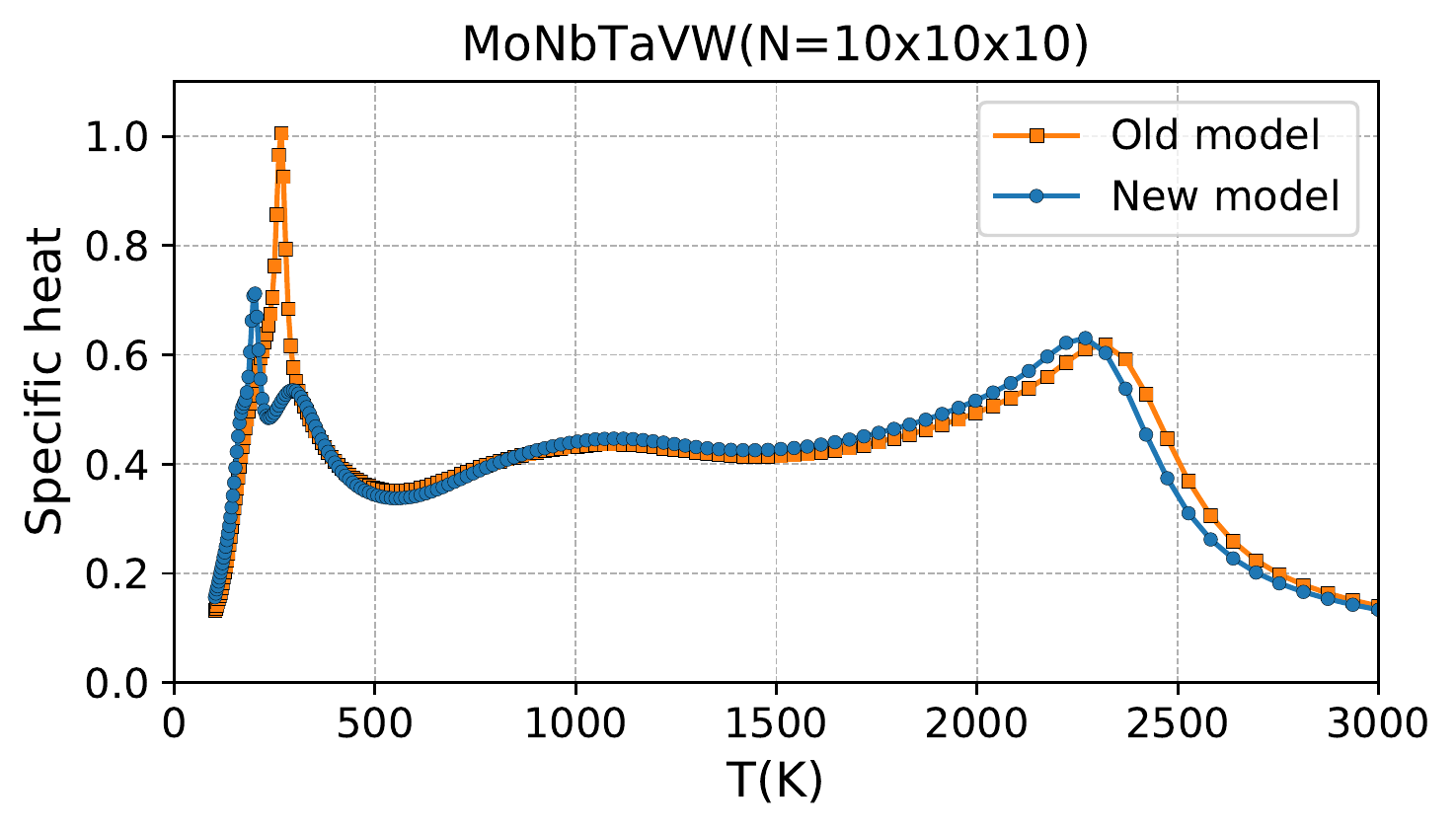} \\
\includegraphics[width=0.45\textwidth]{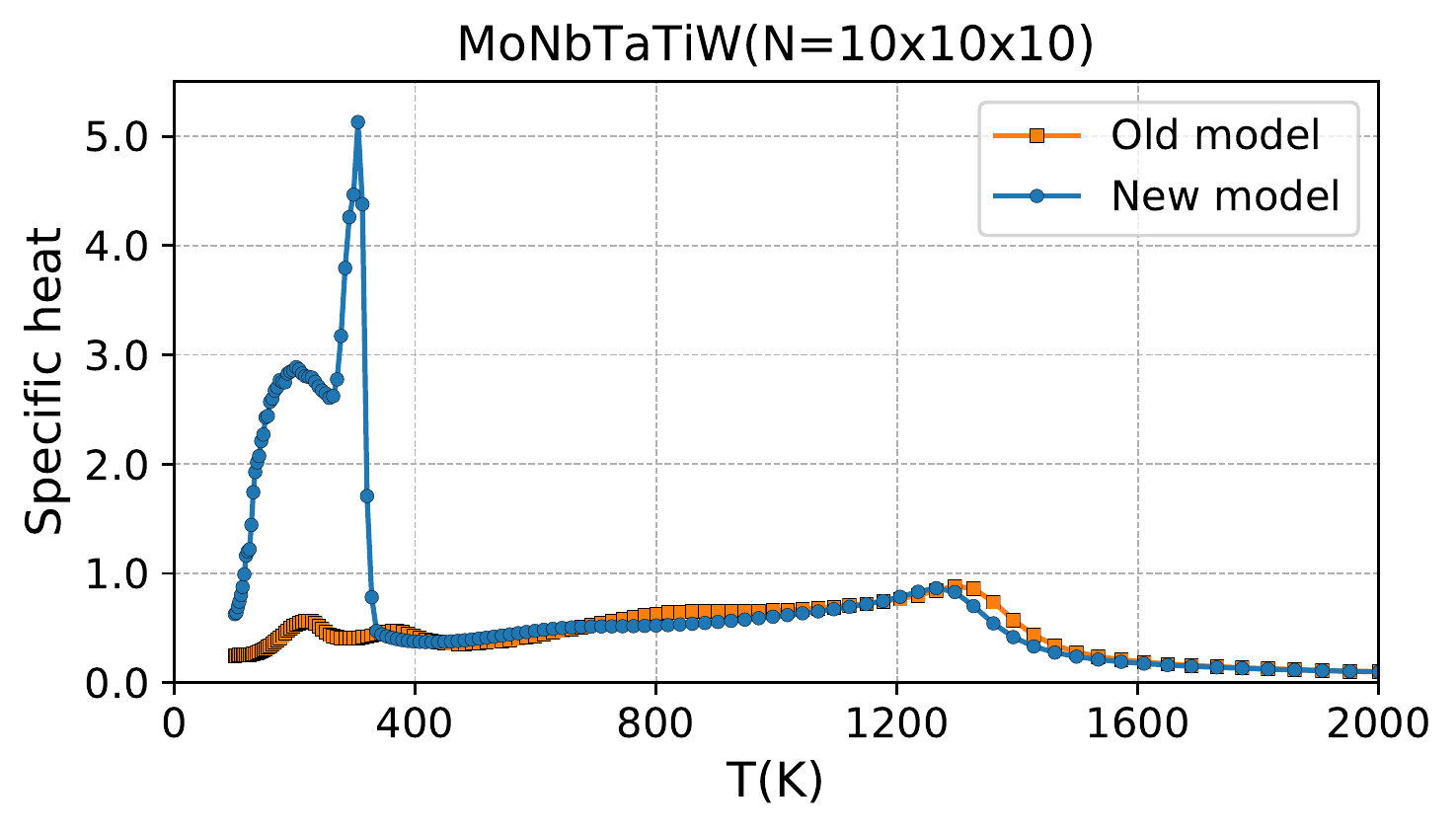}
  \caption{ (Color online) The specific heats of MoNbTaW, MoNbTaVW, and MoNbTaTiW. New model in the legend means the EPI model is fitted with both random and MC configurations, while the old model represents using only the random configurations.} 
\label{fig:specific_heat}
\end{figure}

\begin{figure}[!ht]    
    \centering
\includegraphics[width=0.4\textwidth]{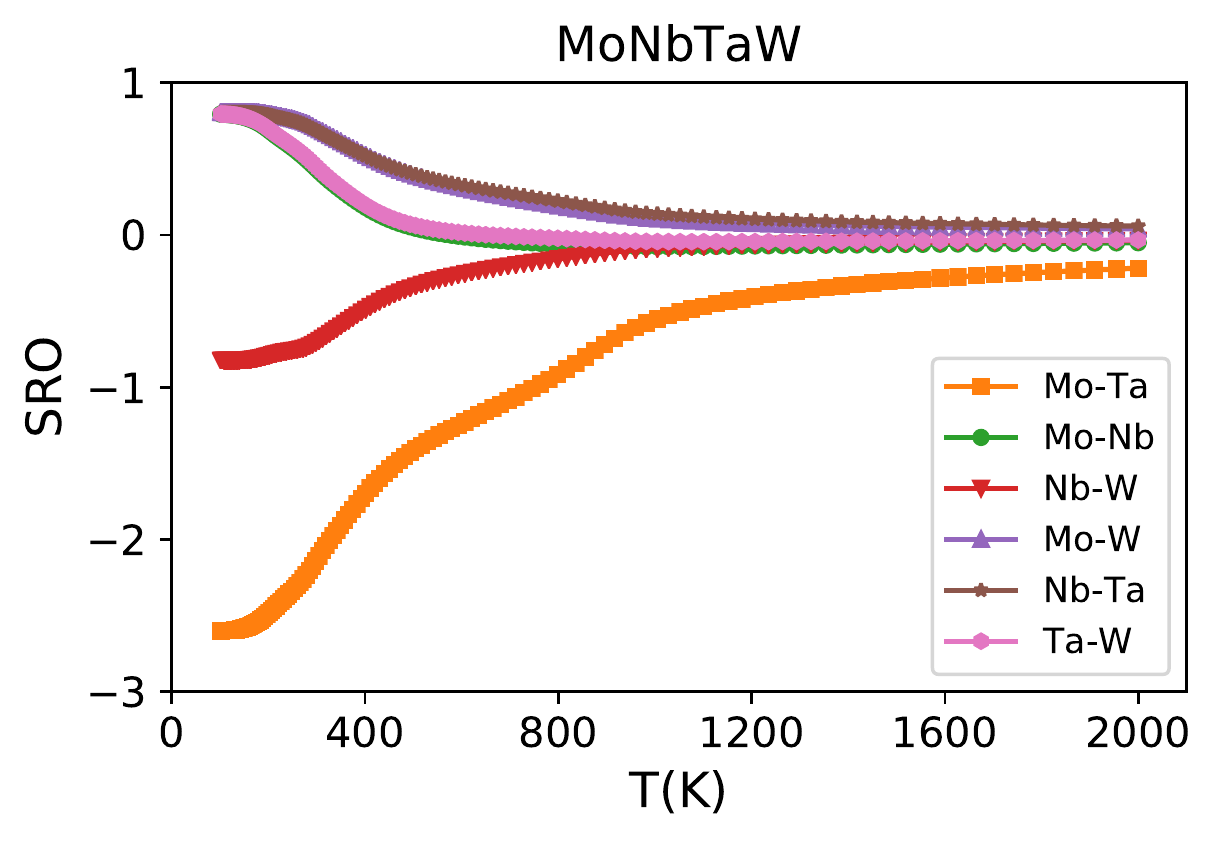}
\includegraphics[width=0.4\textwidth]{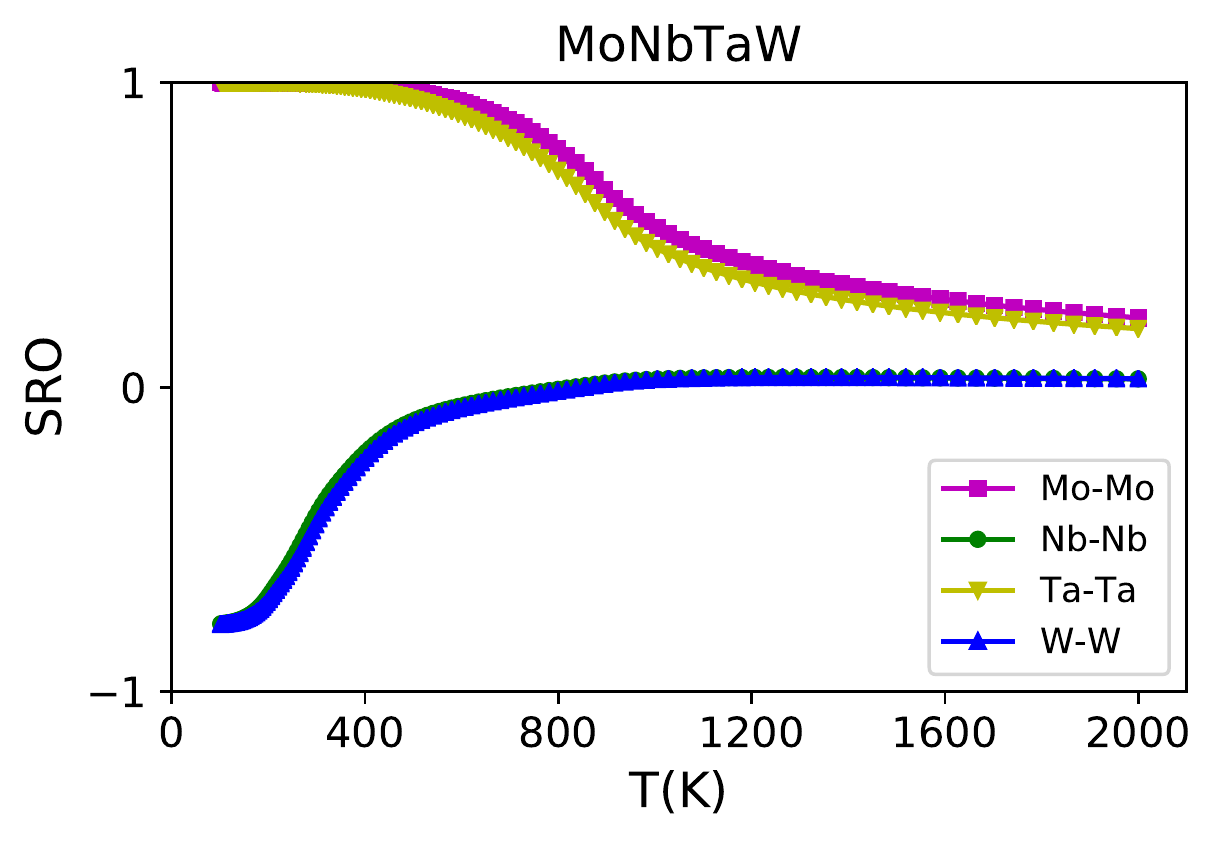}
\includegraphics[width=0.4\textwidth]{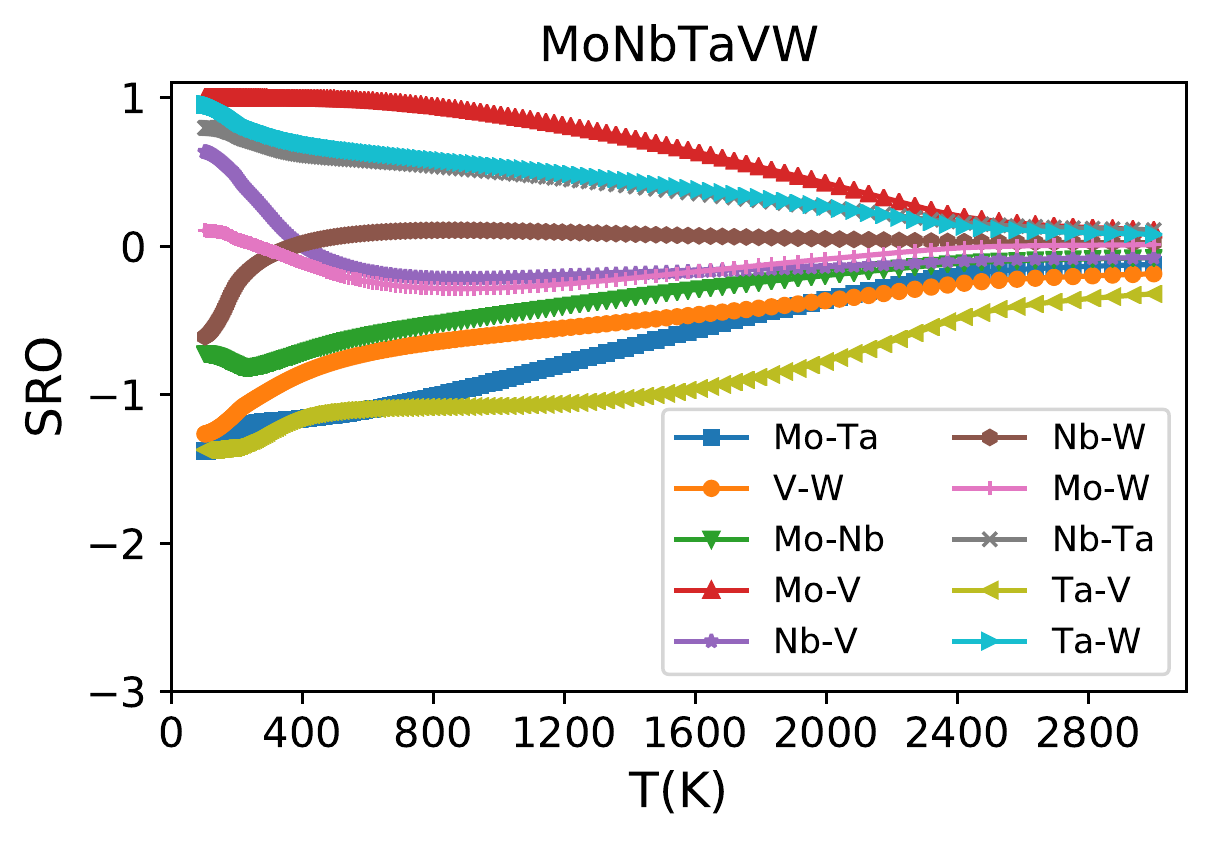}
\includegraphics[width=0.4\textwidth]{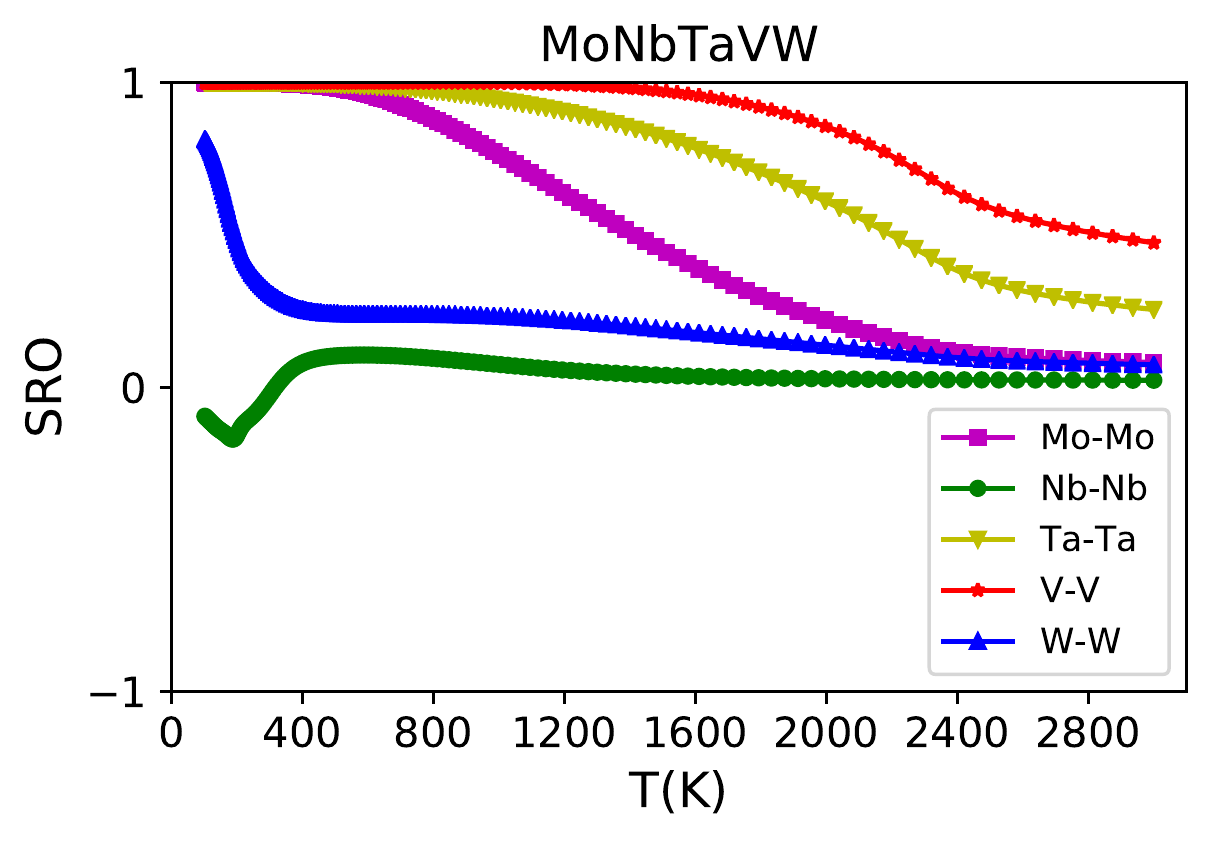}
\includegraphics[width=0.4\textwidth]{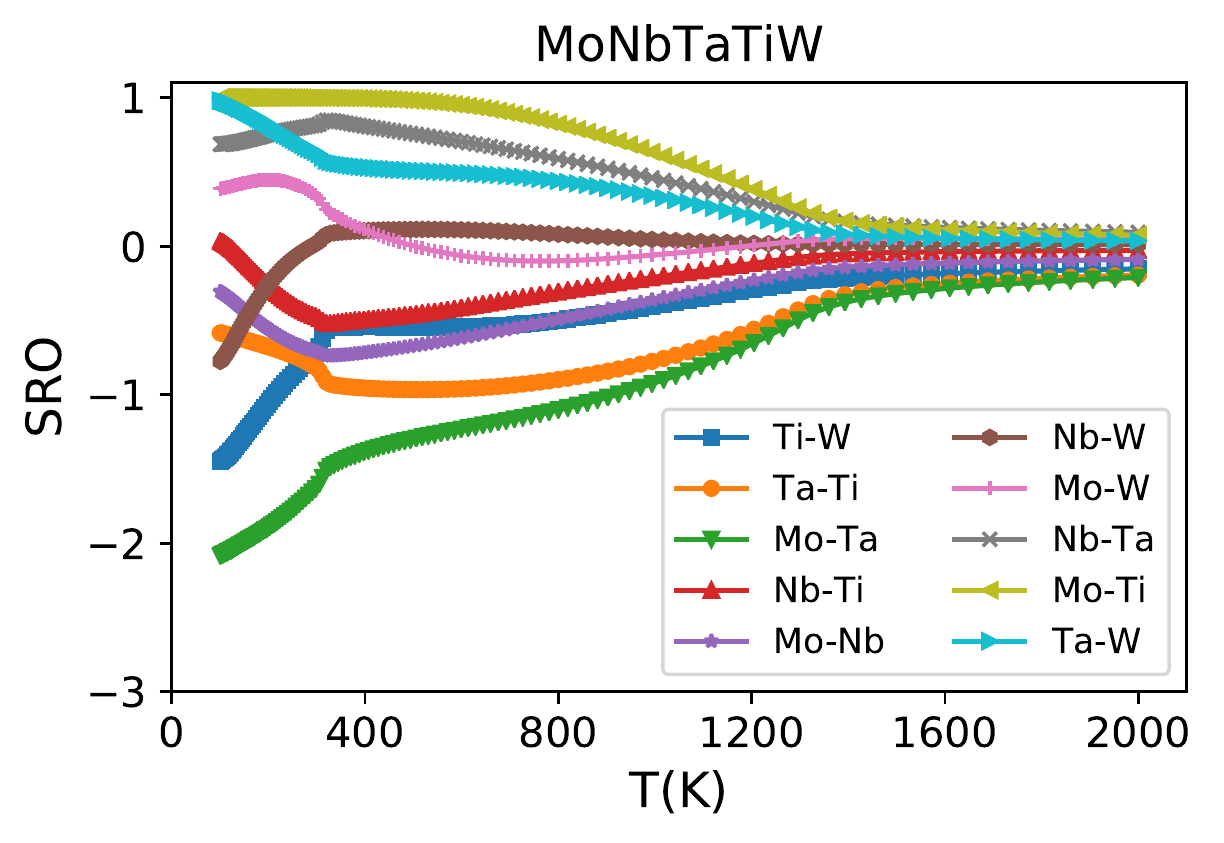}
\includegraphics[width=0.4\textwidth]{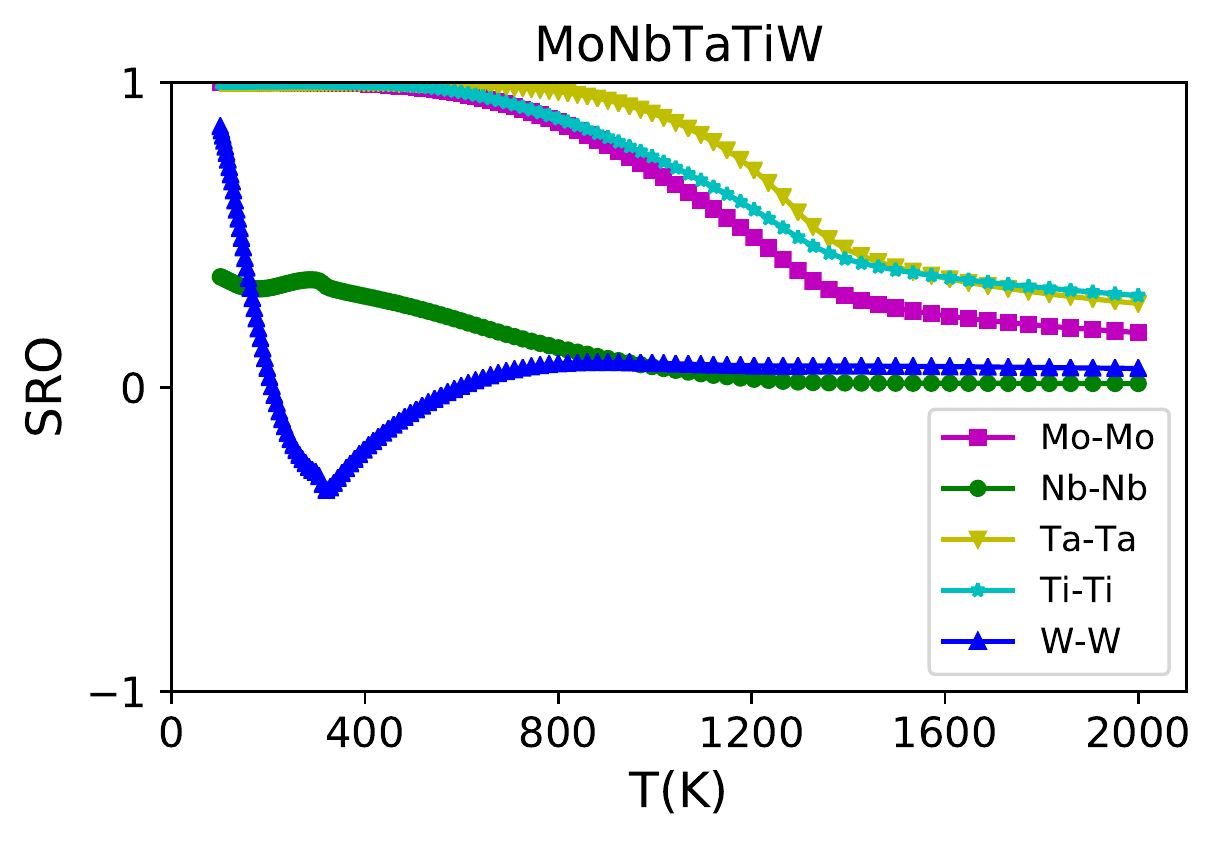}
  \caption{ (Color online) The nearest-neighbor short-range order parameters of MoNbTaW, MoNbTaVW, and MoNbTaTiW. The left side plots are for pairs of different elements, while the right side ones are for pairs of the same element. } 
\label{fig:SRO}
\end{figure}

The thermodynamics of MoNbTaW has been well studied in other works \cite{Huhn2013, e18080403, Korman_npj}. For test purposes, the chemical configurations of MoNbTaW at three different temperatures are shown in Fig.~\ref{fig:snap}. It is not difficult to see that at 100 K, MoNbTaW segregates into two phases, with Mo and Ta forms B2 structures.  When the temperature increases to 304 K, some Nb and W atoms move into the Mo-Ta phase, and the B2 structure is partially broken. At $T = 2000 K$, the ordered structure vanishes completely and the system forms a random solid solution. These results are in agreement with the calculation results in literature \cite{Huhn2013, e18080403, Korman_npj}. To investigate the thermodynamics of the refractory HEAs in detail, the specific heats $C_v$ and the nearest neighbor (NN) SRO parameters at different temperatures are calculated from canonical Monte Carlo and shown in Fig.~\ref{fig:specific_heat} and \ref{fig:SRO}. The specific heats from the EPI model using only the random data are also shown in Fig.~\ref{fig:specific_heat} as a comparison. Note that although the improvement of the model is only about 0.1 mRy, it still has a significant impact on the curve, especially at low temperature and phase transition point. The result further highlights the importance of using the MC data to improve the energy model.

The most noticeable feature in Fig.~\ref{fig:specific_heat} is that there are mainly two phase-transition peaks in the $C_v$ curve. The first transition occurs near room temperature at $T_1$, and the second one occurs at an elevated temperature of $T_2$. The origin of the phase transitions can be better understood from the SRO parameters of the same elements, as shown in the right side of Fig.~\ref{fig:SRO}. It can be seen that for all the three materials, $T_1$ is due to an order-disorder transition of Nb and W, while $T_2$ is due to the order-disorder transition of the other elements. In other words, it is easier for W and Nb to form random phase, compared to the other elements. From the left side of Fig.~\ref{fig:SRO}, it is easy to see that the Nb-W SRO parameter indeed quickly drops to zero as temperature increases. Moreover, this is also consistent with the results in Fig.~\ref{fig:EPI}, where the pair interactions involving Nb and W tend to be weak. As a result, W and Nb tend to occupy lattice sites randomly in the material rather than form ordered compounds. These observations provide a possible explanation to the excellent combined mechanical properties of HEAs: {\color{black}the random phases provide good ductility, while the ordered precipitates enhance the strength by impeding the movement of dislocations. Similar mechanisms are also reported in other works \cite{Ding8919, LIU2019107955, NatureComm_Tailor}. }

\begin{figure}[!ht]    
    \centering
\includegraphics[width=0.45\textwidth]{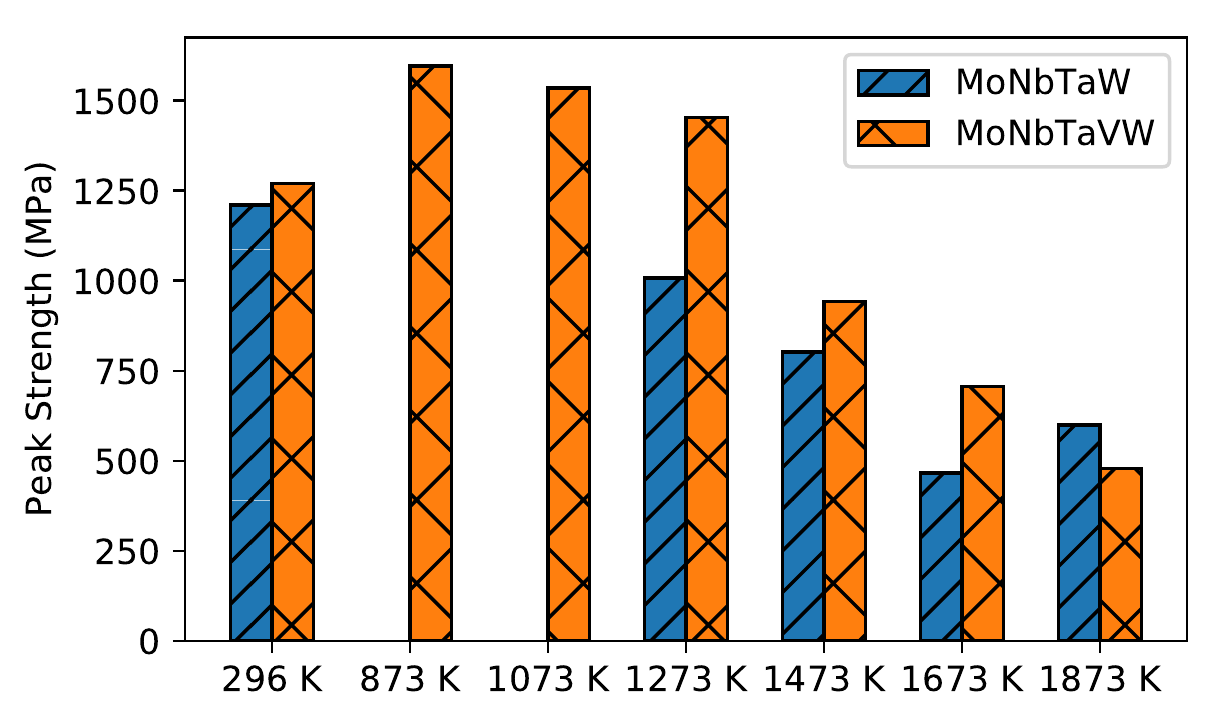} 
\includegraphics[width=0.45\textwidth]{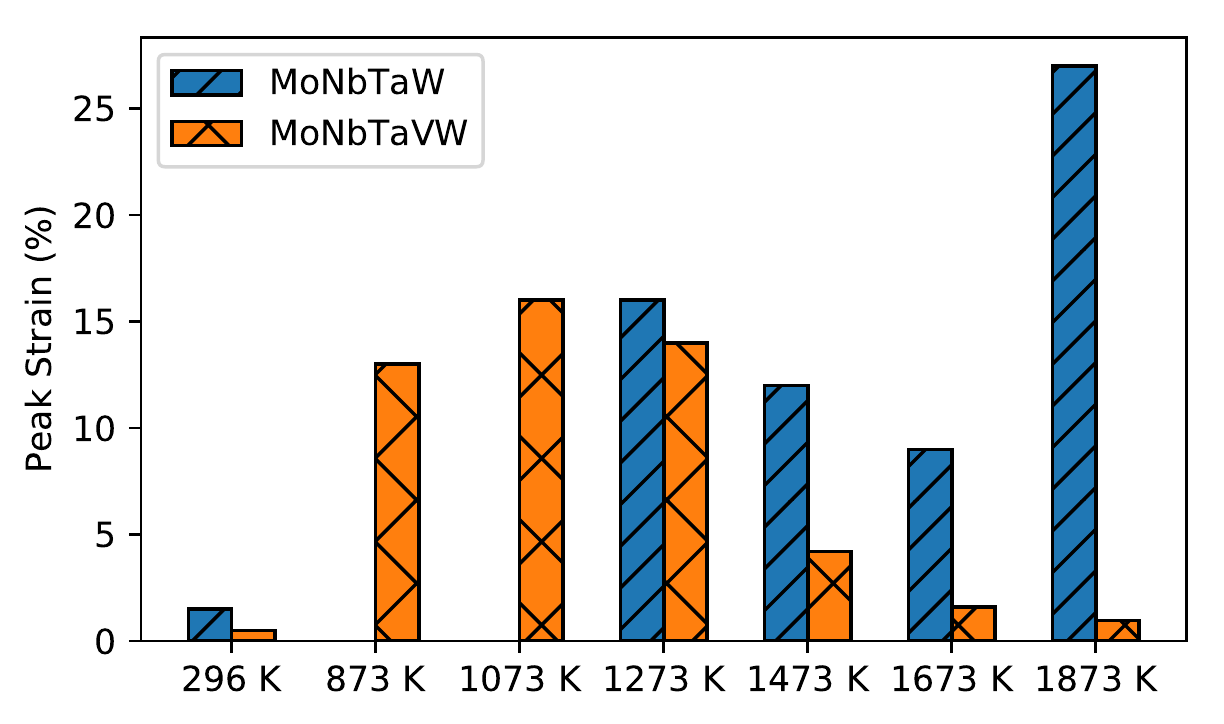} 
  \caption{ (Color online) The peak strength and peak strain for MoNbTaW and MoNbTaVW at different temperatures. The data is from Ref.~\cite{SENKOV2011698}. The values of MoNbTaW at 873 K and 1073 K are not measured.} 
\label{fig:strength_strain}
\end{figure}

{\color{black} This theory is supported by the results shown in Fig.~\ref{fig:specific_heat}.} It can be seen that that the addition of V drastically enhances the value of $T_2$ in MoNbTaVW, which is about 2300 K, as compared to 870 K for MoNbTaW and 1300 K for MoNbTaTiW. As discussed in section \ref{EPI}, this is expected because V demonstrates much stronger bonding with other elements. This strong bonding favors the formation of ordered compounds. As a result, even the order-disorder transition temperature of Ta and Mo has been increased, as shown in Fig.~\ref{fig:SRO}. Therefore, compared to MoNbTaW and MoNbTaTiW, we expect MoNbTaVW to demonstrate better strength, but poorer ductility, {\color{black} due to the large amount of ordered precipitates or second phases. The distinction of the materials is particularly prominent} in the temperature range between 1300 K and 2000 K, where MoNbTaW and MoNbTaTiW are predominantly random solid solutions, while MoNbTaVW is still largely ordered. This is in excellent agreement with the experimental stress-strain curve in Ref.~\cite{SENKOV2011698}. For convenience, the peak strength and peak strain data in Ref.~\cite{SENKOV2011698} are shown in Fig~\ref{fig:strength_strain}. We also expect to see an increase of ductility among all the materials around $T_1$ (about 300--400 K), due to the order-disorder transition of W and Nb. This is also in agreement with the result in Ref.~\cite{SENKOV2011698}, where the fracture strain of MoNbTaVW is found to be only 1.7 \% at 296 K, but quickly increase to 13 \% when measured at 873 K, as shown in Fig~\ref{fig:strength_strain}. {\color{black} Of course, the strength and ductility of a material depends on the various factors at different length scales, therefore more detailed study is required to draw a more quantitative conclusion.}

\section{Conclusion}
We develop a novel data-driven framework to construct effective Hamiltonian for the study of thermodynamics in HEAs using a data-driven approach. Compared to traditional DFT methods, the use of the linear-scaling LSMS greatly improves the calculation speed, allowing the use of a relatively larger DFT dataset and the direct evaluation of the configurations from Monte Carlo simulation. By using the effective pair interactions as the features and adopting the regularized Bayesian regression, the learned effective Hamiltonians demonstrate excellent robustness. By systematically adding data from the Monte Carlo samples, the representativeness of the datasets is greatly improved and the obtained effective Hamiltonian demonstrates very high predicted accuracy, with the test RMSEs as small as 0.019, 0.044, and 0.099 mRy respectively for MoNbTaW, MoNbTaVW, and MoNbTaTiW. These small errors are particularly critical for the study of low temperature phases and order-disorder transition in thermodynamics. 

Using the learned effective Hamiltonian, we investigate the evolution of the specific heats and short-range order parameters through canonical Monte Carlo simulation. For all the studied materials, we demonstrate that there are two major order-disorder transitions, one occurring near room temperature and another one at a higher temperature. We identify that the first transition is caused by W and Nb, while the second one is due to the other elements. We conclude that these results provide an explanation for the stress-strain relations found in the experiment. For example, the addition of V introduces strong pair interactions, which significantly increases the temperature of the second order-disorder transition. { \color{black} As a result, the abundance of second-phase precipitates in a wide temperature range} reduces the ductility of the MoNbTaVW, as compared to MoNbTaW and MoNbTaTiW. Moreover, the first order-disorder transition in the materials also helps explain the experimental phenomenon of ductility increase after room temperature. These findings will provide useful guidance and insight to the future design of HEAs.  

{\color{black} We would also like to make some comments about our results. First, the high accuracy of the EPI Hamiltonian demonstrates that for fixed chemical concentrations, the pair interactions are the dominant ones in the investigated materials. We think this is quite reasonable considering that the elements in HEAs generally have similar chemical properties. Of course, whether this generally holds true still requires further investigations on other HEA systems. Second, while the EPI model is highly accurate for canonical Monte Carlo simulation, it is interesting to ask how the pair interactions would be affected if the chemical concentrations change. From Fig. 5, it is easy to see that for different materials, the magnitude of the EPIs of the same chemical pairs are very close, which indicates that the EPIs are not sensitive to the chemical concentrations. On the other hand, the change of the chemical concentrations will generally affect the number of valence electrons, the Fermi surface, and the electronic structure, which may require more complicated form of effective Hamiltonians to describe. For such a case, advanced machine learning methods may perform better than a simple linear regression. }


{ \color{black} Finally, we want to provide some perspective on the computational cost. Note that most of the computational time is spent on the DFT calculation. Compared to the conventional method that directly calls DFT calculations at each Monte Carlo (MC) step, our proposed method only needs a much smaller number of DFT calculations for training and then constructs an efficient data-driven EPI model as a surrogate for MC simulations. In particular, we use 200 DFT data for EPI model training with additional 50 DFT data from Monte Carlo samples to improve the data representativeness. Although a little piece of time is spent on data processing and training, the proposed data-driven method is still much less than the conventional method, which typically needs more than millions of DFT calculations. In this case, an approximate estimation is that the computational cost of the data-driven method is less than 0.0025\% of the brutal force method with simple directly combing DFT and MCs. In particular, from the perspective of CPU core hours cost, the approximately 50 DFT energies from the Monte Carlo sample are the most expensive ones, with each energy data taking about 200 CPU core hours, so the total computational cost is about 10,000 core hours. This is easy for current supercomputers, while the conventional statistical simulation method mentioned in the introduction requires more than millions of core hours. Moreover, if one only uses a smaller 100-atom supercell for the DFT data, then the computational cost of one single data point is reduced to about 20 core hours. This means the calculation can be carried out on regular workstation computers. For a 100-atom supercell, it is also practical to include lattice relaxation effects using other common DFT implementations (e.g., pseudopotential planewaves, linearized augmented planewaves), at the price of longer computational time than LSMS due to their cubic scaling behaviors.}

\section{Data availability statement}
The data of this study will be made available on request.

\section{Acknowledgements}
The work of X. L. and M. E. were supported by the U.S. Department of Energy, Office of Science, Basic Energy Sciences, Materials Science and Engineering Division. The work of J. Z. was supported by the U.S. Department of Energy, Office of Science, Office of Advanced Scientific Computing Research, Applied Mathematics program under contract ERKJ352; and by the Artificial Intelligence Initiative at the Oak Ridge National Laboratory (ORNL). The work of J. Y. was supported by the U.S. Department of Energy, Office of Science, National Center for Computational Sciences. The work of Y.W. was supported in part by NSF Office of Advanced Cyberinfrastructure and the Division of Materials Research within the NSF Directorate of Mathematical and Physical Sciences under award number 1931525. This research used resources of the Oak Ridge Leadership Computing Facility, which is supported by the Office of Science of the U.S. Department of Energy under Contract No. DE-AC05-00OR22725.

\clearpage

 \bibliographystyle{unsrt} 
\bibliography{sample.bib}

\end{spacing}
\end{document}